\documentclass[]{aa}

\usepackage{graphicx}
\usepackage{txfonts}
\usepackage{xcolor}
\usepackage{comment}
\newcommand{\bl}[1]{\mbox{\boldmath$ #1 $}}
\begin{document}

   \title{Formation of pebbles in (gravito-)viscous protoplanetary disks with various turbulent strengths}


   \author{ Eduard I. Vorobyov\inst{1,2},
           Vardan G. Elbakyan
          \inst{3,2},
          Anders Johansen\inst{4,5}, Michiel Lambrechts\inst{4,5}, Aleksandr M. Skliarevskii\inst{2}, and Stoyanovskaya, O. P.\inst{6,7}
          }
   \institute{ University of Vienna, Department of Astrophysics, T\"urkenschanzstrasse 17, 1180, Vienna, Austria; 
   \email{eduard.vorobiev@univie.ac.at} 
   \and
     Research Institute of Physics, Southern Federal University, Rostov-on-Don 344090, Russia   
   \and
     School of Physics, University of Leicester, Leicester, LE1 7RH, UK  
        \and
            Centre for Star and Planet Formation, Globe Institute, University of Copenhagen, {\O}ster Voldgade 5–7, 1350 Copenhagen, Denmark
    \and
   Lund Observatory, Department of Astronomy and Theoretical Physics, Lund University, Box 43, 22100 Lund, Sweden            
        \and
Lavrentyev Institute of Hydrodynamics SB RAS, 15 Lavrentyev Ave., 630090 Novosibirsk, Russia
\and
Mechanics and Mathematics Department, Novosibirsk State University, 2 Pirogov str., 630090 Novosibirsk, Russia
}

\date{}

   
   \titlerunning{Formation of pebbles in protoplanetary disks}
   \authorrunning{Vorobyov et al.}

  \abstract
  {}
   {Dust plays a crucial role in the evolution of protoplanetary disks. We study the dynamics and growth of initially sub-$\mu m$ dust particles in
   self-gravitating young protoplanetary disks with various strengths of turbulent viscosity. We aim to understand the physical conditions that determine the formation and spatial distribution of pebbles when both disk self-gravity and turbulent viscosity can be concurrently at work.}
   {We perform the thin-disk hydrodynamics simulations of self-gravitating protoplanetary disks over an initial time period of 0.5~Myr using the FEOSAD code. Turbulent viscosity is parameterized in terms of the spatially and temporally constant $\alpha$-parameter, while the effects of gravitational instability on dust growth is accounted for by calculating the effective parameter $\alpha_{\rm GI}$. We consider the evolution of dust component including  momentum exchange with gas, dust self-gravity, and also a simplified model of dust growth.}
   {We find that the level of turbulent viscosity strongly affects the spatial distribution and total mass of pebbles in the disk.  The $\alpha=10^{-2}$ model is viscosity-dominated, pebbles are completely absent, and dust-to-gas mass ratio deviates from the reference 1:100 value no more than by 30\% throughout the disk extent. On the contrary, the $\alpha=10^{-3}$ model and, especially, the $\alpha=10^{-4}$ model are dominated by gravitational instability. The effective parameter $\alpha+\alpha_{\rm GI}$ is now a strongly varying function of radial distance. As a consequence, a bottle neck effect develops in the innermost disk regions, which makes gas and dust accumulate in a ring-like structure. Pebbles are abundant in these models, although their total mass and spatial extent is sensitive to the dust fragmentation velocity and to the strength of gravitoturbulence. The use of the standard dust-to-gas mass conversion is not suitable for estimating the mass of pebbles. }
   {Our numerical experiments demonstrate that pebbles can be abundant in protoplanetary disks already at the initial stages of disk evolution. Dust growth models that consider disk self-gravity and ice mantles may be important for studying planet formation via pebble accretion.}


   \maketitle
%


\section{Introduction}

A fundamental problem of the planet formation theory is how the micron-sized grains coagulate and grow into km-sized planetesimals and later into planets. One of the obstacles here is known as the ``radial drift'' problem -- the inward radial motion of dust grains caused by friction with gas on timescales shorter than a protoplanetary disk lifetime
\citep{1972Whipple, 1976Adachi, 1977Weidenschilling}. A promising solution to this problem is the streaming instability \citep{2005YoudinGoodman}, which leads to the formation of dense clumps of solid particles \citep{2011Johansen, 2014YangJohansen}, which compactify into solid  objects with a few hundreds of km in size by the action of self-gravity \citep{2012Johansen}. The dynamics of such large objects is no longer affected by gas and hence by rapid inward migration. 

Another solution for the radial drift problem is the trapping of dust grains in the substructures of gaseous disks, such as gaseous clumps forming in gravitationally unstable disks \citep{1998Boss, 2017Nayakshin, 2019VorobyovElbakyan} or local pressure maxima also known as radial pressure bumps \citep{1972Whipple, 2003HaghighipourBoss, 2009Johansen}. Ring structures that are frequently observed in protoplanetary disks \citep{2018LongPinilla, 2019vanderMarel} may be associated with the radial pressure bumps \citep{2019Perez}. Several mechanisms that can form radial pressure bumps in different parts of the disk include the disk–planet interaction \citep{2012ZhuNelson, 2015Dipierro, 2018DongLi}, zonal flows caused by the magnetorotational instability \citep[hereafter, MRI;][]{2009Johansen}, and radial variations in the disk density and/or viscosity caused by the radially varying MRI strength  \citep{2007KretkeLin,2012Pinilla,2016Pinilla,2018DullemondPenzlin,2019Charnoz}. Another possible location for a radial pressure bump is the innermost disk region, where a transition from the MRI-active to the MRI-dead zone may occur \citep{2010Dzyurkevich,  2019Ueda, 2017Flock, 2019Flock}.  Anticyclonic vortices at sharp viscosity transitions can also create local pressure maxima \citep{2009LyraJohansen,2012Regaly}.

Radial pressure bumps act not only as a stopping barrier for inward-drifting dust, but also serves as an effective dust growth environment. Dust accumulation in the radial pressure bump increases the local dust-to-gas mass ratio and can lead to the planetesimal formation via the streaming instability \citep{2005YoudinGoodman, 2007JohansenYoudin} or via the gravitational instability \citep{1981Coradini, 2014ChatterjeeTan} in combination with the pebble accretion \citep{2019LambrechtsMorbidelli, 2019IzidoroBitsch, 2020Morbidelli}.

The magnitude of turbulence in the disk is one of the key parameters that impacts the mass and angular momentum transport \citep{1974LyndenBell}, orbital evolution of planets \citep{2012Kley, 2011Paardekooper}, and evolution of dust in the disk  \citep{2007OrmelCuzzi, 2012Birnstiel,2018VorobyovAkimkin}. Turbulence caused by the MRI is usually parametrized through a dimensionless $\alpha$-parameter \citep{1973ShakuraSunyaev} in hydrodynamic models that do not simulate the MRI explicitly. Despite a number of attempts, turbulence in a protoplanetary disk is hard to measure directly from observations \citep{2016Teague, 2017Flaherty}. A wide range of $\alpha$-values from $10^{-4}$ to 0.1 were found in different observational studies \citep{2012MuldersDominik, 2016Pinte, 2018AnsdellWilliams, 2018Dullemond, 2020Flaherty, 2020Rosotti}. Numerical magnetohydrodynamics simulations of the MRI are also not conclusive, yielding values of the $\alpha$-parameter from 0.1--0.01 for fully MRI-active disks \citep{2018Yang,2020Zhu} to $10^{-4}$ when non-ideal magnetohydrodynamics effects are taken into account \citep{2015Bai,2018Simon}.  

Self-gravity is another major player in disk evolution, at least in its early stages \citep{2014Turner,2016Kratter}. Gravitational torques can dominate the viscous ones in young protoplanetary disks \citep{VorobyovBasu2009}, gas and dust dynamics can be affected by gravitational instability \citep{2004RiceLodato,2020Riols}, and even the longevity of vortices can be reduced owing to gravitational torques \citep{2017RegalyVorobyov}. The effects of gravitational instability may also be expressed in terms of an effective viscous parameter $\alpha_{\rm GI}$ \citep{2008Kratter, 2010Vorobyov_visc, 2016Kratter, 2018RiolsLatter}, making it easier to assess its effect on disk dynamics and dust growth.

The main focus of the present paper is on studying how the magnitude of turbulent viscosity caused by the MRI and gravitoturbulence caused by gravitational instability  influences the dynamics and growth of dust in young protoplanetary disks. To make things as simple as possible, we consider the viscous $\alpha$-parameter to be a constant of time and space, with its value corresponding to either fully MRI-active ($10^{-2}$) or MRI-reduced ($10^{-3}$) or  MRI-suppressed ($10^{-4}$) disk. In addition, we investigate how gravitational instability and gravitoturbulence can work together with turbulent viscosity to shape the gas and dust disks in their early stages of evolution.

In particular, we are interested in the spatial distribution and total mass of pebbles as a key ingredient in the pebble accretion model of planet formation.  For this purpose, we use the FEOSAD numerical hydrodynamics code \citep{2018VorobyovAkimkin}, which allows us to study the formation and long-term evolution of circumstellar disks starting from the prestellar core collapse phase. The evolution of the disk is considered self-consistently inside a gravitationally contracting envelope, which serves as a reservoir of gas and small (micron-sized) dust particles in the early embedded stages of disk evolution. A simple model of dust growth is employed and various dust fragmentation velocities are considered to assess their effect on the spatial distribution of pebbles in the disk.

This paper is organized as follows. In Sect.~\ref{sec:model} we describe our numerical model, paying emphasis to new features as compared to \citet{2018VorobyovAkimkin}. In Sect.~\ref{sec:results} we present our main results. Results for the parameter space study are presented in Sects.~\ref{sec:param} and \ref{sec:vert_turb}. 
Total masses of pebbles are presented in Sect.~\ref{sect:masses}.
We finally draw our conclusions in Sect.~\ref{sec:concl}. In the Appendix, we present a semi-analytic explanation of the "bottle neck" effect in our models, compare the radial drift velocities of dust, and describe the Stokes regime of dust dynamics in the innermost disk regions.

\section{Numerical model}\label{sec:model}
The formation and evolution of a protoplanetary gas-dust disk is studied using the FEOSAD  (Formation and Evolution Of Stars and Disks) two-dimensional numerical hydrodynamics code. The code is described in detail in \citet{2018VorobyovAkimkin} and here we briefly review the key features of the code and its subsequent improvements.

Our numerical simulations start from the gravitation collapse of a rotating flattened prestellar core and proceed through the protostar and disk formation phases. The simulations are terminated in the T~Tauri phase of disk evolution when the age of the system reaches 0.5 Myr. 
When compared to one-dimensional viscous disk models \citep[e.g.,][]{2016Kimura,2017Drazkowska}, our numerical model is advantageous as it can follow more realistically the formation and evolution of non-axisymmetric structures and dust drift in the disk. The thin-disk model also has its shortcomings compared to fully three-dimensional models \citep[e.g.,][]{2019Desai,2020Zhu}, namely, the vertical motions are neglected and the local hydrostatic equilibrium is imposed. We note that the adopted thin-disk limit is different from the razor-thin approximation in the sense that the vertical scale height of the disk is calculated using the assumption of local hydrostatic equilibrium in the gravitational field of both star and disk. This quantity is further used in the calculation of the fraction of stellar irradiation absorbed by the disk surface and in the computations of disk characteristics related to dust drift. The properties of the central protostar are calculated using the stellar evolution tracks obtained with the STELLAR code \citep{2008YorkeBodenheimer,2013HosokawaYorke}. The stellar mass grows according to the mass accretion rate from the disk, and the radiative heating of the disk is calculated in accordance with the protostellar photospheric and accretion luminosities.

The numerical model takes into account the viscous and shock heating, irradiation from the central star and from the circumstellar background environment, dust radiative cooling from the disk surface, momentum exchange between gas and dust (including backreaction of dust on gas), self-gravity of gas and dust disks, and turbulent viscosity using the $\alpha$-parametrization \citep{1973ShakuraSunyaev}. For the simulations, we use a two-dimensional polar grid ($r,\phi$) with $400\times256$ grid zones. The radial grid is spaced logarithmically, while the azimuthal grid is distributed uniformly. The chosen relation between the number of grid zones in the radial and azimuthal directions produces grid cells with a square-like shape, thus minimizing numerical errors when computing the fluxes.   

To avoid prohibitively small time-steps on the converging two-dimensional polar grid, we replace the innermost 0.2~au region of the disk with a sink cell. We emphasize that the size of the sink cell in our simulations is notably smaller than in many other (including our own) global disks simulations over time scales of hundreds of kyr.  The use of the thin-disk approximation makes possible long integration times with such a small sink cell. The inner boundary condition is carefully chosen to avoid the development of an artificial density drop near the disk-sink interface by allowing matter to flow in both directions across the inner boundary \citep[see][for details]{2018VorobyovAkimkin}. 
The free outflow condition is imposed on the outer boundary so that the matter is allowed to flow out of the computational domain, but is prevented from flowing in.

\subsection{Gas component}
The FEOSAD code considers the co-evolution of gas and dust disk subsystems. Both the gas and dust components are modelled as fluids. The dust is treated as a pressureless fluid.
The hydrodynamic equations of mass, momentum, and energy transport for the gas component are as follows
\begin{equation}
\label{cont}
\frac{{\partial \Sigma_{\rm g} }}{{\partial t}} +  \nabla  \cdot ( \Sigma_{\rm g} {\bl v} ) = 0,  
\end{equation}
\begin{equation}
\label{mom}
\frac{\partial}{\partial t} \left( \Sigma_{\rm g} {\bl v} \right) +  \nabla \cdot \left(\Sigma_{\rm g} {\bl v} \otimes {\bl v} \right) = - \nabla {\cal P}  + \Sigma_{\rm g} \, {\bl g}  + \nabla \cdot \mathbf{\Pi}  - \Sigma_{\rm d,gr} {\bl f},
\end{equation}
\begin{equation}
\label{energ}
\frac{\partial e}{\partial t} +\nabla \cdot \left( e {\bl v} \right) = -{\cal P} (\nabla \cdot {\bl v}) -\Lambda +\Gamma + \left(\nabla {\bl v}\right):\Pi, 
\end{equation}
where $\Sigma_{\rm g}$ is the gas surface density, $\Sigma_{\rm d,gr}$ is the grown dust surface density described in more detail later in this section, $e$ is the internal energy per surface area,  ${\cal P}$ is the vertically integrated gas pressure calculated via the ideal equation of state as ${\cal P}=(\gamma-1) e$ with $\gamma$=7/5, ${\bl v}=v_r \hat{\bl r}+ v_\phi \hat{\boldsymbol \phi}$  is the gas velocity in the disk plane,  $\nabla=\hat{\bl r} \partial / \partial r + \hat{\boldsymbol \phi} r^{-1} \partial / \partial \phi $ is the gradient along the planar coordinates of the disk, ${\bl g}=g_r \hat{\bl r} +g_\phi \hat{\boldsymbol \phi}$ is the gravitational acceleration in the disk plane due to the gravity of central protostar and self-gravity of gas and dust in the disk \citep{2010VorobyovBasu}, $\mathbf{\Pi}$ is the viscous stress tensor, the expression for which can be found in \citet{2010VorobyovBasu}. The kinematic viscosity is expressed following the \citet{1973ShakuraSunyaev} ansatz   $\nu = \alpha c_{\mathrm{s}} H_{\rm g}$, where $c_\mathrm{s}$ and  $H_{\rm g}$ are the sound speed and gas vertical scale height, respectively. The $\alpha$-parameter is set to spatially and temporally constant values: $10^{-2}$, $10^{-3}$, and $10^{-4}$. The terms $\Lambda$ and $\Gamma$ are, respectively, the cooling and heating rates, the expressions for which can be found in \citet{2018VorobyovAkimkin}. The cooling rate takes into account the blackbody cooling from the surface of the disk, while the heating rate accounts for the heating due to the stellar and background irradiation. The dust opacities are taken from \citet{2003SemenovHenning}. We emphasize, however, that we do not apply the typical 1:100 dust-to-gas scaling when calculating the disk optical depth and use the total dust column densities directly derived from numerical modeling. The form of the friction force $\bl f$ is provided in the next section.

\begin{figure}
\begin{centering}
\includegraphics[width=1\columnwidth]{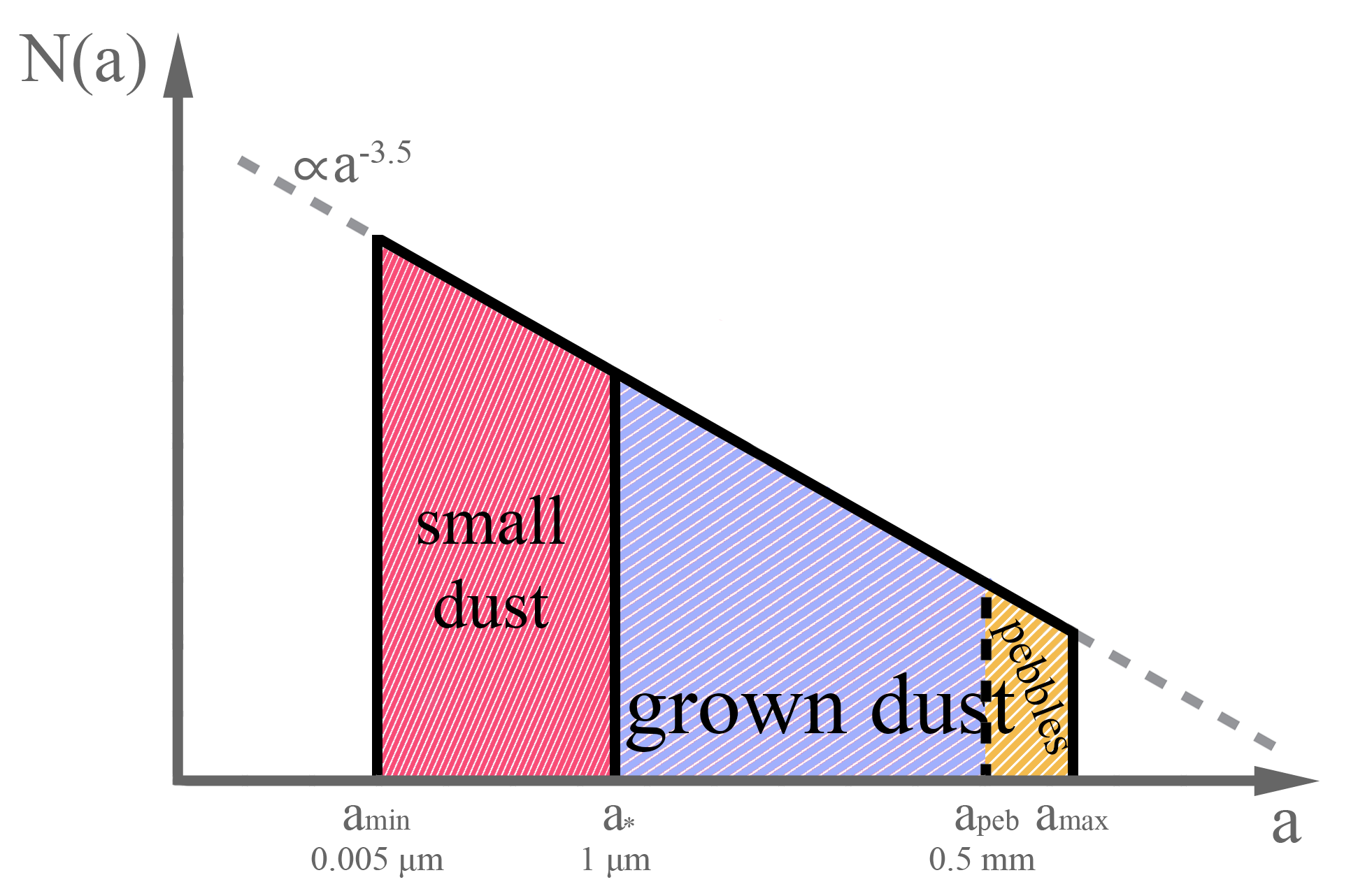}
\par\end{centering}
\caption{\label{fig:0} Illustration of the dust size distribution in our model. The red area represents the amount of small dust, the blue and orange areas together represent the amount of grown dust, while only the orange area represents the amount of pebbles.}
\end{figure}

\subsection{Dust component}
The dust component in our model consists of small sub-micron dust and grown dust. The  size distribution of both dust populations in our model, $dN/da = C a^{-p}$, has a fixed power law of $p$=3.5 with a normalization constant $C$ and is schematically depicted in Figure~\ref{fig:0}. The minimum size\footnote{When referring to the size of a dust particle, we mean its radius.} of small dust particles in the model is $a_{\rm min}$=$0.005~\mu\rm m$ and the maximum size is $a_*$=$1~\mu\rm m$. For the grown dust, $a_*$ is the fixed minimum size and  $a_{\rm max}$ is the variable maximum size.
Initially, only small (sub-micron) dust exists in the collapsing prestellar core, but small dust can grow and transform into grown dust as the disk forms and evolves. In the FEOSAD code small dust is assumed to be dynamically coupled to the gas, while the dynamics of grown dust is controlled by friction with the gas and by the total gravitational potential of the system. The effect of dust-to-gas friction on both gas and dust velocities is taken into account using the analytic integration method, which  belongs to a wider set of asymptotic preserving methods \citep[e.g.][]{2017Stoyanovskaya}. Good performance of the chosen integration scheme on the standard Sod and dusty wave test problems was demonstrated in \citet{2018Stoyanovskaya}.

The continuity and momentum equations for small and grown dust components are defined as 
\begin{equation}
\label{contDsmall}
\frac{{\partial \Sigma_{\rm d,sm} }}{{\partial t}}  + \nabla  \cdot \left( \Sigma_{\rm d,sm} {\bl v} \right) = - S(a_{\rm max}),  
\end{equation}
\begin{equation}
\label{contDlarge}
\frac{{\partial \Sigma_{\rm d,gr} }}{{\partial t}}  + \nabla  \cdot \left( \Sigma_{\rm d,gr} \, {\bl u} \right) =  \nabla \cdot \left( D \Sigma_{\rm g} \nabla \left( {\Sigma_{\rm d,gr} \over \Sigma_{\rm g}} \right)  \right) + S(a_{\rm max}),  
\end{equation}
\begin{equation}
\label{momDlarge}
\frac{\partial}{\partial t} \left( \Sigma_{\rm d,gr} \, {\bl u} \right) +  \nabla \cdot \left( \Sigma_{\rm d,gr} \, {\bl u} \otimes {\bl u} \right)   =  \Sigma_{\rm d,gr} \, {\bl g} + \Sigma_{\rm d,gr} {\bl f} + S(a_{\rm max}) {\bl v},
\end{equation}
where $\Sigma_{\rm d,sm}$ and $\Sigma_{\rm d,gr}$ are the surface densities of small and grown dust, $\bl u= u_r \hat{\bl r}+ u_\phi \hat{\boldsymbol \phi}$ describes the planar components of the grown dust velocity and $D$ is the turbulent diffusivity of grown dust, which is related to the kinematic viscosity as $D=\nu / \mathrm{Sc}$ \citep{1988ClarkePringle}. The Schmidt number $\mathrm{Sc}$ is taken to be unity in this study. 
The term $\bl f$ is the drag force per unit dust mass between dust and gas, which is defined as
\begin{equation}
    {\bl f} = {1\over 2 m_{\rm d}} C_{\rm D} \, \sigma \rho_{\rm g} ({\bl v} - {\bl u}) |{\bl v} - {\bl u}|,
\end{equation}
where $\sigma=\pi a^2$ is the cross-section of dust grains, $\rho_{\rm g}$ the gas volume density, $m_{\rm d}$ the mass of a dust grain, and $C_{\rm D}$ the dimensionless friction parameter. The latter quantity is usually defined using the approximation formula of \citet{1977Weidenschilling}. Assuming further that the mean free path of molecular hydrogen $\lambda$ is much greater than the size of a dust grain $a$ (the Epstein regime), the friction coefficient can be written as $C_{\rm D}= 8 c_{\rm s}/(3 |{\bl v} - {\bl u}|)$, so that the friction force can be conveniently expressed as
\begin{equation}
  \label{eq:friction_force}
    {\bl f }= \frac{{\bl v} - {\bl u}}{t_{\rm stop}},
\end{equation}
where $t_{\rm stop}$ is the stopping time 
\begin{equation}
 \label{tstop}
    t_{\rm stop} = {\rho_{\rm s} a \over \rho_{\rm g}c_{\rm s} },
\end{equation}
where $\rho_{\rm s} = 3 m_{\rm d} / (4 \pi a^3) =  2.24$~g~cm$^{-3}$ is the material density of dust grains.

We found, however, that in the innermost disk regions ($r\la 1.0$~au) the Epstein regime may be violated because the mean free path of hydrogen molecules becomes shorter than the size of dust particles (the Stokes regime, see Appendix~\ref{Stokes}). \citet{1977Weidenschilling} provides the approximation formulae for $C_{\rm D}$ in the Stokes regime as well. However, \citet{2020Stoyanovskaya} demonstrated that the Weidenschilling approach becomes inaccurate in the transonic regime and for large Reynolds numbers $\mathrm{Re}=4 \, a \, \mathrm{Ma}/\lambda$, where $\mathrm{Ma}=|{\bl v} - {\bl u}|/ c_{\rm s}$ is the Mach number.
Therefore, we re-defined the stopping time using the friction coefficient from \citet{1976Henderson}, which can be written in terms of the Mach and Reynolds numbers for $\mathrm{Ma}<1.0$ as
\begin{eqnarray}
    C_{\rm D}&=&\displaystyle\frac{24}{\mathrm{Re}+S \left( 4.33+\displaystyle 1.567 \exp(-0.247 \frac{\rm{Re}}{S})\right)} \nonumber \\
    &+& 0.6 S\left(1-\exp\left(-\displaystyle\frac{\rm Ma}{\rm Re}\right)  \right) \nonumber \\
    &+&\exp\left(-0.5\displaystyle\frac{\rm Ma}{\sqrt{\rm Re}}\right) \left( \displaystyle
    \frac{4.5+0.38 \left(0.03 {\rm Re}+0.48 \sqrt{\rm Re}\right)}{1+0.03 {\rm Re} + 0.48 \sqrt{\rm Re} } \right. \nonumber  \\ 
    &+& \left. 0.1 {\rm Ma}^2+0.2 {\rm Ma}^8 \right),
\end{eqnarray}
and for $\mathrm{Ma}>1.75$ as
\begin{equation}
    C_{\rm D}=\displaystyle\frac{0.9+\displaystyle \frac{0.34}{\rm{Ma}^2}+1.86\sqrt{\displaystyle \frac{\rm Ma}{\rm Re}}\left(2+\displaystyle\frac{2}{S^2}+\displaystyle\frac{1.058}{S}-\displaystyle\frac{1}{S^4}\right)}{1+1.86\displaystyle\sqrt{\frac{\rm Ma}{\rm Re}} },
\end{equation}
where $S=\mathrm{Ma}\sqrt{\gamma/2}$. The intermediate values for the drag coefficient ($1<\mathrm{Ma}<1.75$) can be obtained using a linear interpolation \citep{2020Stoyanovskaya}.
Evidently, the use of the Henderson formulae comes at the expense of higher computational costs, but it was shown to behave well in the flow regimes where the Weidenschilling approximation fails \citep{2020Stoyanovskaya}.

The stopping time for the Henderson drag coefficient is defined as
\begin{equation}
t_{\rm stop} = {8 \over 3} {a \rho_{\rm s} \over \rho_{\rm g} C_{\rm D} |{\bl v} - {\bl u}|  }.
\end{equation}
Here, the gas volume density is found as $\rho_{\rm g}=\Sigma_{\rm g}/(\sqrt{2\pi} H_{\rm g})$. When calculating the stopping time, we used $a_{\rm max}$ as a characteristic dust size rather than a mean value obtained by averaging over the entire dust size distribution of the grown dust population (i.e., from $a_\ast$ to $a_{\rm max}$). 
This approach is justified, because the main subject of this study is the dynamics of pebbles, which have sizes close to the $a_{\rm max}$ value.

\subsection{Small to grown dust conversion}
The term $S(a_{\rm max})$ is the conversion rate of small dust into the grown dust per unit surface area, which can be expressed as
\begin{equation}
    S(a_{\rm max}) = -\frac{\Delta\Sigma_{\mathrm{d,sm}}}{\Delta t},
    \label{growth:rate}
\end{equation}
where $\Delta\Sigma_{\mathrm{d,sm}}=\Sigma_{\mathrm{d,sm}}^{n+1}- \Sigma_{\mathrm{d,sm}}^{n}$ is the mass of small dust (per surface area $\Delta S$) converted to grown dust during one hydrodynamic time step $\Delta t$.
For the chosen dust size distribution ($dN/da = C a^{-p}$) the masses of small and grown dust per surface area $\Delta S$ at the beginning of the time step ($\Sigma_{\mathrm{d,sm}}^n$ and $\Sigma_{\mathrm{d,gr}}^n$) and at the end of the time step ($\Sigma_{\mathrm{d,sm}}^{n+1}$ and $\Sigma_{\mathrm{d,gr}}^{n+1}$) can be expressed as
\begin{equation}
    \Sigma_{\mathrm{d,sm}}^n = \frac{4\pi \rho_s}{3\Delta S} C_{\mathrm{sm}}^n \int_{a_{\mathrm{min}}}^{a_*}a^{3-\mathrm{p}}da, \;
    \Sigma_{\mathrm{d,gr}}^n = \frac{4\pi \rho_s}{3\Delta S} C_{\mathrm{gr}}^n \int_{a_*}^{a_{\mathrm{max}}^{n}}a^{3-\mathrm{p}}da,
\end{equation}
\begin{equation}
    \Sigma_{\mathrm{d,sm}}^{n+1} = \frac{4\pi \rho_s}{3\Delta S} C_{\mathrm{sm}}^{n+1} \int_{a_{\mathrm{min}}}^{a_*}a^{3-\mathrm{p}}da, \;
    \Sigma_{\mathrm{d,gr}}^{n+1} = \frac{4\pi \rho_s}{3\Delta S} C_{\mathrm{gr}}^{n+1} \int_{a_*}^{a_{\mathrm{max}}^{n+1}}a^{3-\mathrm{p}}da,
\end{equation}
where $C_{\mathrm{sm}}^{n}$ and $C_{\mathrm{gr}}^{n}$ are the normalization constants for the small and grown dust at the beginning of the time step, while $C_{\mathrm{sm}}^{n+1}$ and $C_{\mathrm{gr}}^{n+1}$ are the corresponding quantities at the end of the time step.
Since the total mass of dust ($\Sigma_{\mathrm{d,tot}}^n=\Sigma_{\mathrm{d,sm}}^n+\Sigma_{\mathrm{d,gr}}^n=\Sigma_{\mathrm{d,tot}}^{n+1}$) in a specific grid cell does not change in the process of dust growth,
the surface density of small dust at the beginning of the time step ($\Sigma_{\mathrm{d,sm}}^n$) and at the end of the time step ($\Sigma_{\mathrm{d,sm}}^{n+1}$) can be presented as
\begin{equation}
    \Sigma_{\mathrm{d,sm}}^n = \Sigma_{\mathrm{d,tot}} \frac{ C_{\mathrm{sm}}^n I_1}{C_\mathrm{sm}^n I_1 + C_{\mathrm{gr}}^n I_2}, \;  \Sigma_{\mathrm{d,sm}}^{n+1} = \Sigma_{\mathrm{d,tot}} \frac{ C_{\mathrm{sm}}^{n+1} I_1}{C_\mathrm{sm}^{n+1} I_1 + C_{\mathrm{gr}}^{n+1} I_3},
\end{equation}
where
\begin{equation}
I_1 = \int_{a_{\mathrm{min}}}^{a_*}a^{3-\mathrm{p}}da, \; I_2 = \int_{a_*}^{a_{\mathrm{max}}^{n}} a^{3-\mathrm{p}}da, \; I_3 = \int_{a_*}^{a_{\mathrm{max}}^{n+1}} a^{3-\mathrm{p}}da.
\end{equation}

To calculate $\Sigma_{\mathrm{d,sm}}^n$ and $\Sigma_{\mathrm{d,sm}}^{n+1}$, and hence the conversion rate of small to grown dust given by Equation~(\ref{growth:rate}), the normalization constants have to be determined. In our earlier study \citep{2020Elbakyan}, we assumed that 
$C_{\mathrm{sm}}^{n} = C_{\mathrm{gr}}^{n}$ and $C_{\mathrm{sm}}^{n+1} = C_{\mathrm{gr}}^{n+1}$, effectively implying that no discontinuity in the dust size distribution appears at $a_*$ as the disk evolves with time (see Fig.~\ref{fig:0}). However, due to different drift timescales of small and grown dust particles, the surface densities of small and grown dust can change in such a manner that a discontinuity could develop in the dust size distribution at $a_\ast$.

To account for this effect in the present study, we assume that the normalization constants $C_{\mathrm{sm}}^{n}$ and$ C_{\mathrm{gr}}^{n}$ are generally distinct, while the normalization constants $C_{\mathrm{sm}}^{n+1}$ and $C_{\mathrm{gr}}^{n+1}$ are set to be equal to each other. This effectively corresponds to the assumption that dust growth smooths out any discontinuity in the dust size distribution at $a_\ast$ that may appear due to differential drift of small and grown dust populations.
With this assumption, the amount of small dust $\Delta\Sigma_{\mathrm{d,sm}}$ (per surface area $\Delta S$) converted to grown dust during one hydrodynamic time step $\Delta t$ (instead of Eq. (12) from \citet{2018VorobyovAkimkin}) is calculated as

\begin{equation}
\label{deltasigma}
\begin{split}
    \Delta\Sigma_{\mathrm{d,sm}} =  \Sigma_{\mathrm{d,sm}}^{n+1} - \Sigma_{\mathrm{d,sm}}^n 
  = \Sigma_{\mathrm{d,tot}} \frac{I_1\left( C_{\mathrm{gr}}^n I_2 - C_{\mathrm{sm}}^n I_3\right)}{I_4\left( C_{\mathrm{sm}}^{n} I_1 + C_{\mathrm{gr}}^{n} I_2\right)},
\end{split}
\end{equation}
where
\begin{equation*}
\label{c1}
    C_{\mathrm{sm}}^{n} = 
    \frac
    {
    3 \Sigma_{\mathrm{sm}}^n \Delta S
    }
    {
    4 \pi \rho _{s} I_1
    }, \;
    C_{\mathrm{gr}}^{n} = 
    \frac
    {
    3 \Sigma_{\mathrm{gr}}^n \Delta S
    }
    {
    4 \pi \rho _{s} I_2
    }, \;
    I_4 = I_1+I_3 = \int_{a_{\mathrm{min}}}^{a_{\rm max}^{n+1}} a^{3-\mathrm{p}}da.
\end{equation*}
Substituting $C_{\mathrm{sm}}^{\rm n}$ and $C_{\mathrm{gr}}^{\rm n}$ into Eq.~(\ref{deltasigma}) and assuming a conservation of total dust mass, we finally obtain
\begin{equation}
\label{deltasigmafinal}
   S(a_{\rm max})= -\frac{\Delta\Sigma_{\mathrm{d,sm}}}{\Delta t}= {1\over \Delta t} \left( {\Sigma_{\mathrm{d,sm}}^n } -
    \frac
    {
    \Sigma_{\mathrm{d,tot}}
    I_1  
    }
    {
    I_4
    } \right).
\end{equation}

For the chosen slope of the dust size distribution $p=3.5$, the ratio ${I_1}/{I_4}$ is equal to  
    $(\sqrt{a_*} - \sqrt{a_\mathrm{min}}) /  
    (\sqrt{a_\mathrm{max}} - \sqrt{a_\mathrm{min}})$, meaning that the conversion rate $S(a_{\rm max})$ is inverse proportional to $\sqrt{a_{\rm max}}$ and decreases as dust grows.
    
To complete the calculation of $S(a_{\rm max})$, the maximum size  of  grown  dust $a_{\rm max}$ in  a  given  computational  cell must be computed at each time step using the following equation
\begin{equation}
\frac{\partial a_{\rm max}}{\partial t} + ({\bl u} \cdot \nabla)a_{\rm max} = \mathcal{D}.
\end{equation}

The second term on the left-hand side describes the change of the maximum dust size in a given grid cell due to advection and the source term $\mathcal{D}$ represents the growth rate of dust due to collisional coagulation
\begin{equation}
\mathcal{D} = \frac{\rho_{\mathrm{d}} v_{\mathrm{rel}}}{\rho_{\mathrm{s}}},
\label{growth-rate}
\end{equation}
where $\rho_{\mathrm{d}}$ is the total dust volume density and $v_{\mathrm{rel}}$ is the dust-to-dust collision velocity, which takes the Brownian and turbulent velocity of dust into account. { In particular, the dust-to-dust collision velocity owing to turbulence is computed following the model of turbulent eddies proposed in \citet{2007OrmelCuzzi} 
\begin{equation}
    v_{\rm{turb}} = \sqrt{{3 \alpha \over \mathrm{St}+\mathrm{St}^{-1}}} c_{\rm s},
    \label{turb_vel}
\end{equation}
where $\mathrm{St}$ is the Stokes number. The value of $v_{\rm turb}$ is then added to the velocity of Brownian motions of dust particles to obtain $v_{\mathrm{rel}}$.
}


The total dust volume density is found as 
\begin{equation}
    \rho_{\rm d} = {\Sigma_{\rm d,sm} H_{\rm d} + \Sigma_{\rm d,gr} H_{\rm g} \over \sqrt{2 \pi} H_{\rm d} H_{\rm g} },
    \label{rho:dust}
\end{equation}
where we assumed that the vertical scale height of small dust is equal to that of gas, but grown dust can settle toward the disk midplane, having defined its scale height as a function of the Stokes number $\mathrm{St}$ and $\alpha$-parameter \citep{2004Kornet}.

The dust growth in our model is limited by the so-called fragmentation barrier \citep{2012Birnstiel}. The maximum size up to which dust particles are allowed to grow is defined as
\begin{equation}
\label{afrag}
a_{\rm frag} = \frac{2\Sigma_{\rm g}u^2_{\rm frag}}{3\pi \rho_{\rm s} \alpha c_{\rm s}^2},
\end{equation}
where we choose $u_{\rm frag}$=3~m~s$^{-1}$ as a threshold value for the dust fragmentation velocity \citep{2018Blum}. The effects of varying $u_{\rm frag}$ are discussed in Sect.~\ref{sec:param}.
Thus, when $a_{\rm max}$ exceeds $a_{\rm frag}$, the growth rate $\mathcal{D}$ is set equal to zero and $a_{\rm max}$ is set equal to $a_{\rm frag}$. We note that if the local conditions in the disk change so that $a_{\rm frag}$ drops below the current value of $a_{\rm max}$ (e.g., when temperature increases or gas density decreases), then we also set $a_{\rm max}=a_{\rm frag}$. This effectively implies that part of the grown dust is shattered via collisions and the process of dust conversion reverses, namely, part of grown dust can now be converted to small dust.

\subsection{Definition of pebbles}
The dust dynamics in protoplanetary disks is often characterised by the dimensionless Stokes number, which we define as
\begin{equation}
\mathrm{St}=\frac{\Omega_{\rm K} \rho_{\rm s} a_{\rm max}}{\rho_{\rm g} c_{\rm s}},
\label{eq:stokes}
\end{equation}
where $\Omega_{\rm K}$ is the Keplerian angular velocity and 
$\rho_{\rm g}=\Sigma_{\rm g} / \sqrt{2\pi}H_{\rm g}$ is the gas volume density.
Dust particles with sizes from millimeters to centimeters known as pebbles play a crucial role in the pebble accretion model for planet formation \citep{2010Ormel,2012LambrechtsJohansen, 2016Ida, 2017JohansenLambrechts}. Here, we define pebbles as dust particles that satisfy the following criteria. First, we choose the dust particles with $\mathrm{St}\geq0.01$. This value is widely used as a threshold value for the pebble definition \citep[][]{2012LambrechtsJohansen, 2019LenzKlahr}. Next, using Equation (\ref{eq:stokes}) we find the radius of dust particles $a_{\rm St=0.01}$ at which $\mathrm{St}$ for the local conditions in the disk would be equal to 0.01
\begin{equation}
a_{\rm St=0.01} = a_{\rm max}\frac{0.01}{\mathrm{St}}.
\end{equation} 
If the resulting value of $a_{\rm St=0.01}$ is greater than 0.5~mm, then we define the minimum size of pebbles as $a_{\rm peb,min}=a_{\rm St=0.01}$. Thus, our adopted definition of the minimum pebble size can be expressed as
\begin{equation}
a_{\rm peb,min}=
\begin{cases}
a_{\rm St=0.01}, \, \, \mathrm{if} \, \mathrm{St}\geq0.01 \; \mathrm{and} \; a_{\mathrm{St=0.01}}\geq0.5~\mathrm{mm,} \\
0, \, \, \mathrm{otherwise~(pebbles~do~not~exist).} 
\end{cases}
\label{eq:pebble_def}
\end{equation}
A choice of 0.5~mm as the lower limit on the size of pebbles is motivated by the typical sizes of chondrules, 0.1-1.0~mm \citep{2019Metzler}. Chondrules may have been incorporated to chondrites via the process known as pebble accretion \citep{2015Johansen}. Since pebble accretion is an important mechanism in the planet formation theory, we assume in this work that the minimum size of pebbles roughly corresponds to that of chondrules. We note that for the chosen slope of the dust size distribution ($p=3.5$) the total mass of pebbles is determined by the upper limit on their mass.  

The size distribution of pebbles is schematically illustrated in Figure~\ref{fig:0} with the orange area. We note that $a_{\rm peb,min}$=0  corresponds to the absence of pebbles but grown dust can still be present.  Finally, the surface density of pebbles $\Sigma_{\rm peb}$ inside each computational cell is calculated as
\begin{equation}
\label{peb_mass}
\Sigma_{\rm peb} = \frac{\Sigma_{\rm d,gr} \left(\sqrt{a_{\rm max}} - \sqrt{a_{\rm peb,min}}\right) }{\sqrt{a_{\rm max}} - \sqrt{a_{\ast}}}.
\end{equation}

\subsection{Initial conditions}

Our numerical simulations start from the gravitational collapse of a prestellar core with a mass of $M_{\rm core}=0.59~M_\odot$ and a ratio of rotational-to-gravitational energy of $\beta=2.4 \times 10^{-3}$. Such initial values are consistent with the observations of prestellar cores \citep{2002Caselli} and are chosen to form gravitationally unstable disks that are at the same time stable to fragmentation \citep{2013Vorobyov}.

The radial profiles of gas surface density and angular velocity of the prestellar core are typical for objects with a supercritical mass-to-flux ratio that are formed through ambipolar diffusion, with the specific angular momentum remaining constant during axially-symmetric core collapse \citep{1997Basu}
\begin{equation}
    \Sigma_{\rm g}(r)=\frac{r_0\Sigma_{\rm 0,g}}{\sqrt{r^2+r_0^2}},
\end{equation}
\begin{equation}
    \Omega_{\rm g}(r)=2\Omega_{\rm 0,g}\bigg(\frac{r_0}{r}\bigg)^2\left[\sqrt{1+\left(\frac{r}{r_0}\right)^2}-1\right],
\end{equation}
where $\Sigma_{\rm 0,g}=0.38$~g~cm$^{-2}$ and  $\Omega_{\rm 0,g}=1.8$~km~s$^{-1}$~pc$^{-1}$ are, respectively, the gas surface density and angular velocity at the center of the core, $r_0=620$~au is the radius of the near-uniform central region of the core. The initial gas temperature in the core is 20~K. This value is also set for the temperature of the background disk irradiation.  Initially only small dust is present in the prestellar core, thus the initial surface density of total dust ($\Sigma_{\mathrm{d,tot}}$) is equal to the surface density of small dust ($\Sigma_{\mathrm{d,sm}}$). The surface density of grown dust ($\Sigma_{\mathrm{d,gr}}$), hence the surface density of pebbles ($\Sigma_{\mathrm{peb}}$) initially are equal to zero. The initial total dust-to-gas mass ratio ($\zeta_{\rm d2g}=\Sigma_{\mathrm{d,tot}}/\Sigma_{\mathrm{g}}$) in the prestellar core is equal to 0.01. We note that further in the text we refer to $\zeta_{\rm d2g}$ as dust-to-gas ratio, which must be distinguished from the pebble-to-gas ratio $\zeta_{\rm p2g}=\Sigma_{\mathrm{peb}}/\Sigma_{\mathrm{g}}$, which is initially equal to zero (no pebbles exist in a prestellar core). The pebble-to-gas ratio is always smaller than the dust-to-gas ratio.

\section{Main results}\label{sec:results}
In this section, we present the main results of our numerical simulations, focusing on the disk spatial morphology and the radial distribution of main gas-dust characteristics. We consider three numerical models with different values of the  viscous $\alpha$-parameter equal to $10^{-2}, 10^{-3}$, and $10^{-4}$ but otherwise identical initial characteristics. We note that the $\alpha$-parameter in our models is a constant of time and space.
\subsection{Global disk evolution}

\begin{figure}
\begin{centering}
\includegraphics[width=1\columnwidth]{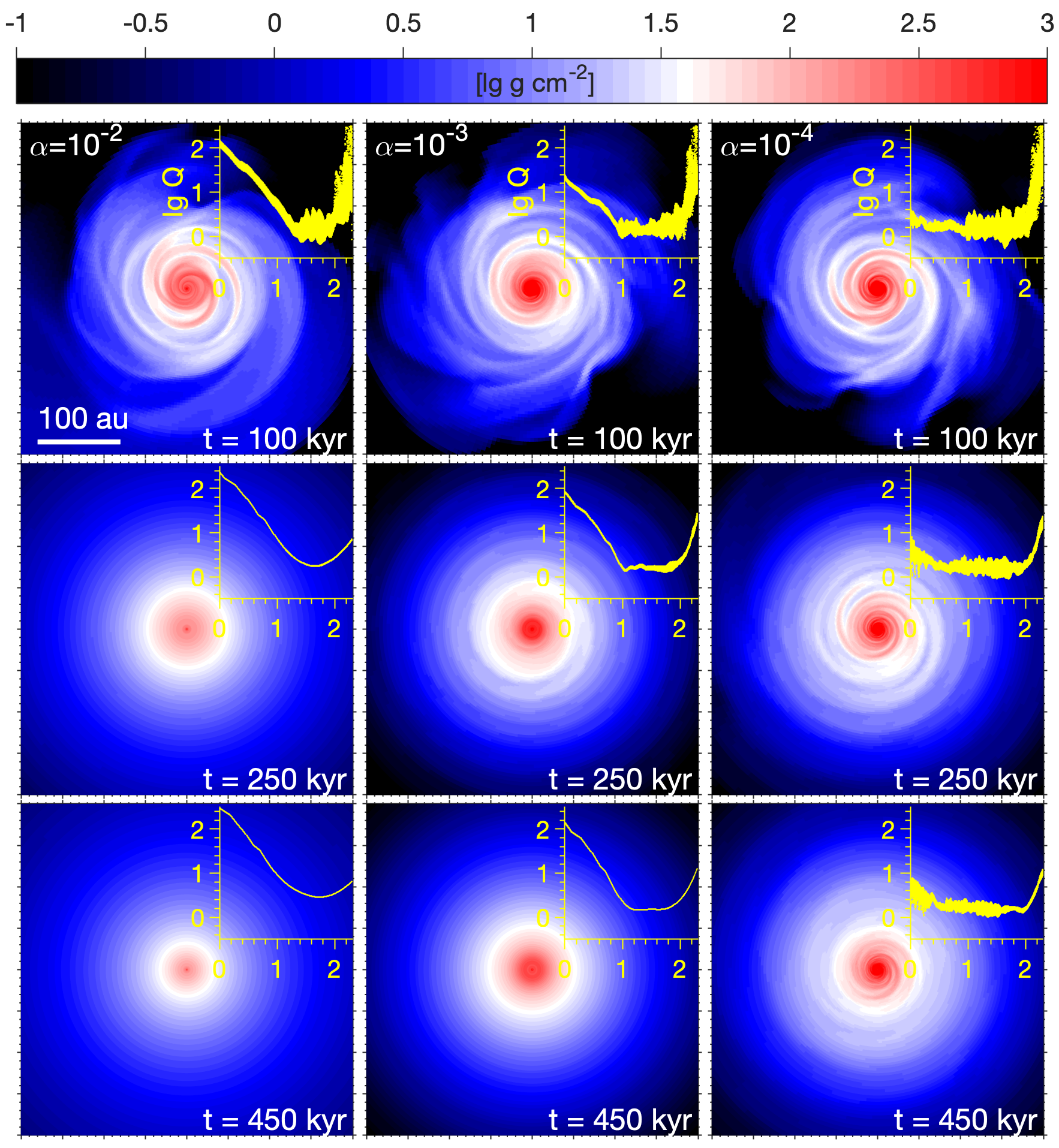}
\par\end{centering}
\caption{\label{fig:1} Temporal evolution of the gas surface density in the inner $400\times400$ au$^2$ box for all three models. The color bar is shown in log scale. The insets show the Toomre $Q$-parameter for all azimuthal grid points at a specific radial distance from the star (both values are in log units).}
\end{figure}

Figure~\ref{fig:1} presents the global disk evolution for all our models over a time period of about 0.5~Myr. The time shown in all figures is counted from the instance of star formation, which occurs $\approx 28$~kyr after the onset of prestellar core collapse. Columns in Figure~\ref{fig:1} show the gas surface density maps corresponding to models with a particular $\alpha$-value. Insets in the upper-right corner of each panel show the Toomre $Q$-parameter \citep{1964Toomre} for all azimuthal grid points at a specific radial distance from the star. The $Q$-parameter  is defined as
\begin{equation}
Q={\frac{\tilde{c}_{\rm s} \Omega} {\pi G (\Sigma_{\rm g}+\Sigma_{\rm d,sm} + \Sigma_{\rm d,gr})}},
\label{eq:toomre}
\end{equation}
where  $\tilde{c}_{\rm s}=c_{\rm s}/\sqrt{1+\zeta_{\rm d2g}}$ is the modified sound speed \citep{2018VorobyovAkimkin} in the presence of dust, $\Omega$ is the angular velocity of gas, and $G$ is the gravitational constant.

During the early evolution, the $Q$-parameter at the radial distances from ten to a few tens of astronomical units in all models drops below a threshold value for the gravitational instability ${Q=1}$ for axisymmetric perturbations \citep{1964Toomre}. We note that for nonaxisymmetric perturbations the disk may become unstable at higher values of $Q$ up to $\sqrt{3}$ \citep{1997Polyachenko}. Spiral arms formed via the gravitational instability in the disk are clearly seen in the top panels of Figure~\ref{fig:1}. During the subsequent evolution, the strength of gravitational instability diminishes, but the rate of this process differs in models with distinct $\alpha$-values. The disk in the $\alpha$=10$^{-2}$ model becomes gravitationally stable already after 0.2~Myr, while the disks in the $\alpha$=10$^{-3}$ and $\alpha=10^{-4}$~models are still unstable at this stage of the evolution.  The disk in the $\alpha$=10$^{-4}$~model stays gravitationally unstable and exhibits a spiral structure during the entire considered evolution period. This difference can be attributed to a more efficient disk viscous spreading in the higher $\alpha$-models. This effect is most pronounced in the $\alpha$=10$^{-2}$ model, the disk of which is characterized by a larger size and lower density compared to the models with lower $\alpha$-parameter. Higher values of $\alpha$ also enhance the mass accretion rate onto the star \citep{VorobyovBasu2009}, which  raises the disk temperature (due to viscous heating) and accelerates the disk mass depletion. Both effects work against gravitational instability, as Equation~(\ref{eq:toomre}) indicates.

\begin{figure*}
\begin{centering}
\includegraphics[width=2\columnwidth]{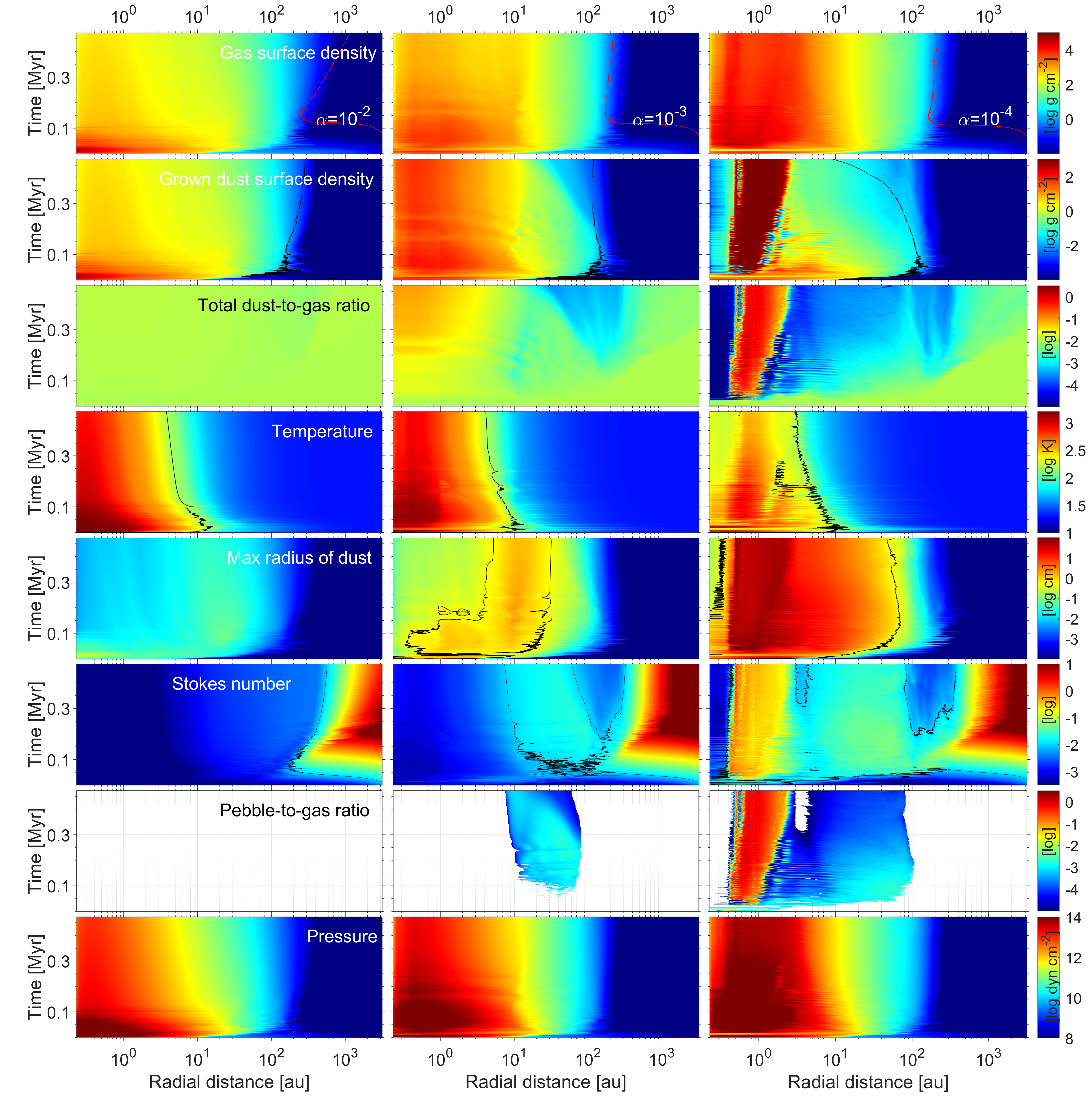}
\par\end{centering}
\caption{\label{fig:2} Temporal evolution of the azimuthally averaged gas surface density (first row), grown dust surface density (second row), total dust-to-gas mass ratio ($\zeta_{\mathrm{d2g}}$) (third row), temperature (fourth row), maximum radius of grown dust (fifth row), Stokes number (sixth row), pebble-to-gas mass ratio ($\zeta_{\mathrm{p2g}}$) (second to bottom row), and gas pressure (bottom row) for models with different $\alpha$-parameter. Color bars are shown in the log scale. The contour lines in the top and second to top rows mark the radial distance $R_{\mathrm{gas}}^{\mathrm{98}}$ and $R_{\mathrm{dust}}^{\mathrm{98}}$, which are the characteristic radial distances containing 98\% of the total (disk+envelope) gas and grown dust, respectively. The contour line in the fourth row shows the radial distance at which the gas temperature is equal to 150 K. The contour line in the fifth row marks the disk regions, where dust size is larger than 0.5~mm. The contour line in the third to bottom  row outlines the region of the disk with $St\geq0.01$.
}
\end{figure*}

\subsection{Azimuthally averaged disk characteristics}
The inner few tens of au of the protoplanetary disk are of a particular interest because this is the region where planets supposedly form. To have a better understanding of the evolution of the inner part of the disk, we present in Figure~\ref{fig:2} the temporal evolution of azimuthally averaged disk characteristics of our models with $\alpha$=10$^{-2}$ (left column), $\alpha$=10$^{-3}$ (middle column), and $\alpha$=10$^{-4}$ (right column). We note that the radial distance in the figure is in the logarithmic scale, which helps to depict the evolution of the inner part of the disk on au and sub-au scales in more details. 

We first consider the $\alpha=10^{-2}$ model. Both gas and grown dust follow a fairly similar evolution pattern, although the spatial distribution of grown dust is slightly more compact than that of gas. The disk during the initial 100~kyr 
is characterized by highest gas and dust densities. This time period of disk evolution is also characterized by intense mass loading from the infalling cloud core, which helps to sustain high densities in the disk. When the core depletes and infall diminishes, the densities of both gas and dust notably decrease. Concurrently, the gas disk begins to  
viscously expand, as can be seen from the $R_{\mathrm{gas}}^{\mathrm{98}}$ curve (the characteristic radial distances containing 98\% of the total gas mass). The spatial expansion of the grown dust component is less pronounced (see the $R_{\mathrm{dust}}^{\mathrm{98}}$ curve), because of inward radial drift. 
We also note that the viscous spreading of the gas disk can cause the mismatch of the gas and dust disk sizes if dust dynamically decouples from gas.
The maximum dust size reaches 0.8~mm, with its radial distribution having a broad peak from several astronomical units to several tens of astronomical units. Interestingly, $a_{\rm max}$ is lower in the innermost disk regions, which can be explained by a decrease in $a_{\rm frag}$ owing to high gas temperatures in these regions.  
The effect of lowering of $a_{\rm frag}$ with rising temperature is evident from Equation~(\ref{afrag}) and is caused by a rising dust-to-dust collision velocity (see Eq.~\ref{turb_vel}).

The gas temperature in the inner 10~au is quite high, initially exceeding 2000~K  in the innermost regions during the first 100~kyr of evolution.  At later times, the disk gradually cools down, mainly owing to a decreasing optical depth caused by an overall decrease in the dust surface density. The black line illustrates the disk gradual cooling by showing the disk loci with $T=150$~K, which roughly correspond to the water snow line, the radial position of which also shrink with time. Nevertheless, the temperatures in the sub-au disk regions remains high after 0.5~Myr of evolution, reaching 700~K.  We note that the dust temperature is equal to that of gas in our models. 

The total dust-to-gas mass ratio deviates from the initial 1:100 value by no more than 30\%, with the mean deviation of just 2\%. Modest deviations from the 1:100 value can be explained by rather small Stokes numbers of grown dust particles throughout the disk extent,  not exceeding 3$\times10^{-3}$ in the inner 100~au. Radial drift velocity of grown dust consists of two components: gradiental $u_{\rm r,grad}$ and advective $u_{\rm r,adv}$ drift velocities. The former can be described by the following analytical approximation \citep{1977Weidenschilling}
\begin{equation}
u_{r,\mathrm{grad}} = - \dfrac{2 V_\mathrm{K} \, \mathrm{St} \, \eta_{\rm dev}}{1 + \mathrm{St}^{2}}, 
\label{u_drift}
\end{equation}
where $V_{\rm K}$ is the Keplerian velocity and $\eta_{\rm dev}$ quantifies the deviation of the gas disk from the Keplerian pattern of rotation and is proportional to pressure gradient in the disk, $d \ln {\cal P} / d \ln r$. The advective drift velocity is \citep{2002TakeuchiLin}
\begin{equation}
    u_{\rm r,adv} = {v_{\rm r} \over 1+ \mathrm{St}^2}.
    \label{u_adv}
\end{equation}
For a steady-state disk the radial component of the gas velocity $v_{\rm r}$ can be written as \citep{1998Hartmann}
\begin{equation}
    v_{\rm r} \simeq {3 \over 2} \alpha c_{\rm s} \left( {H_{\rm g} \over r} \right).
    \label{v_r}
\end{equation}
Equations~(\ref{u_adv}) and (\ref{v_r}) show that the advective drift velocity equal that of gas  if the Stokes number is small, meaning that $u_{\rm r, adv}$ is proportional to the value of $\alpha$-parameter in the disk. In Appendix~\ref{app:vel} we demonstrate that $u_{\rm r,adv}$ exceeds $u_{\rm r,grad}$ in the $\alpha=10^{-2}$ model, which explains rather small deviations of the dust-to-gas ratio from the canonical value in this model.
We note that high Stokes values in the outermost regions ($\sim 10^3$~au) are caused by very small gas surface densities and are of little significance because of lack of grown dust there. 

Pebbles are completely absent in the $\alpha=10^{-2}$ model.
We note that unlike our previous study \citep{2020Elbakyan}, where we used the fragmentation velocity $u_{\rm frag}$=30~m~s$^{-1}$ and the pebbles were present in the models with $\alpha$=10$^{-2}$, here we use the fragmentation velocity $u_{\rm frag}$=3~m~s$^{-1}$, which reduces the maximum size of dust grains below that of pebbles in the $\alpha$=10$^{-2}$~model. 
Recent numerical and experimental studies of silicate dust grains have shown that they stick and grow when their relative collision velocities do not exceed a few m~s$^{-1}$  (see \citet{2018Blum} for a recent review), hence our choice of $u_{\rm frag}$ is more realistic.  The effects of varying $u_{\rm frag}$ are further investigated in Sect.~\ref{sec:param}.

We now turn to the model with a lower value of $\alpha$ set equal to $10^{-3}$. This model shows notable differences from the case with $\alpha=10^{-2}$. 
Both the gas and dust disks have persistent density enhancements in the inner 10~au and are characterized by more compact sizes than in the $\alpha=10^{-2}$ model as can be seen from the $R_{\mathrm{gas}}^{\mathrm{98}}$  and $R_{\mathrm{dust}}^{\mathrm{98}}$ curves.  
Moreover, the grown dust distribution is generally more compact than that of gas, particularly in the late evolution, owing to an increased role of dust drift relative to gas in the low-$\alpha$ disk environment (see Appendix~\ref{app:vel}).  The lasting gas density enhancement in the inner disk is the result of the bottleneck effect, which can occur even in models with spatially constant $\alpha$-parameter because of radially varying gravitational torques in gravitationally unstable protoplanetary disks. We explain this effect in more detail in Appendix~\ref{bottle:neck}. The bottleneck effect causes the development of a local pressure maximum in the inner disk (bottom row in Fig.~\ref{fig:2}), which helps to trap grown dust particles in this part of the disk.

The total dust-to-gas mass ratio $\zeta_{\rm d2g}$ in the disk demonstrates notable deviations from the canonical value, having enhancements in the inner several astronomical units and strong depressions beyond that region. In particular, the dust enhancements relative to the 1:100 value can reach a factor of 9, while the depressions can be as low as a factor of 35. The mean enhancement averaged over the entire gas disk extent (as indicated by the black curves in the top row of Fig.~\ref{fig:1}) is 1.7. The Stokes number of grown dust in the disk of the $\alpha=10^{-3}$ model reaches 2.1$\times10^{-2}$, which implies decoupling of grown dust dynamics from that of gas on considered timescales and explains strong deviations of $\zeta_{d2g}$ from the canonical value of 0.01.

The gas temperature in the innermost regions of the $\alpha=10^{-3}$ model remains to be high, reaching 1500~K in the initial stages of evolution. Although viscous heating is lower in the $\alpha=10^{-2}$ case, dust accumulation increases the optical depth, which somewhat balances the effect of reduced viscous heating.  By the same reason, the optically thick inner disk regions are notably warmer than the rest of the disk. The maximum dust size reaches 2.0~cm in a disk region between several au and 20 au, and is notably higher than in the $\alpha=10^{-2}$ model. The main reason for that is the increased fragmentation barrier in the lower-$\alpha$ model ($a_{\rm frag}$ is inverse proportional to the $\alpha$-parameter, see Eq.~\ref{afrag}). We note that the size of dust particles in the inner few tens of astronomical units is limited by the fragmentation barrier, while in the outer parts it is limited by radial drift \citep{2008BlumWurm, 2017Gonzalez}. Curiously, the maximum size of dust particles drops again in the innermost several au, which is caused by a decrease in $a_{\rm frag}$ due to increased temperature. We therefore find that in both models, $\alpha=10^{-2}$ and $\alpha=10^{-3}$, the maximum dust size is not a monotonic function of radial distance. This may have important consequences for observations of dust disks at mm-wavelengths, leading to the appearance of optical rings that are not associated with physical dust enhancements \citep{2020Akimkin}.

Conditions for pebble formation are now fulfilled in a radial annulus of the disk centered at around 15 au, which shifts closer to the star as the disk evolves. On both sides of this region, either $\mathrm{St}$ or $a_{\rm max}$ or both fall below the values appropriate for pebbles (see Eq.~\ref{eq:pebble_def}). A maximum pebble-to-gas dust mass ratio of $1.5\times10^{-3}$ is reached in the early evolution stage and its value decreases further with time because of efficient inward radial drift. 
We note that pebble formation is initially limited by a radial distance of $r<80$~au, in agreement with the pebble formation zone obtained by \citet{2014LambrechtsJohansen}.

Finally, we consider the model with the lowest value of $\alpha=10^{-4}$ shown in the right column of Figure~\ref{fig:2}. The radial distribution of gas has a strong and persistent density enhancement in the inner 10~au, indicating that the bottleneck effect is the strongest in this model. The radial distribution of grown dust in strikingly different from the corresponding distributions considered in higher-$\alpha$ models. 
While the dust disks in the $\alpha=10^{-2}$ and $\alpha=10^{-3}$ models extend to several tens or even a hundred astronomical units (see the $R_{\mathrm{dust}}^{\mathrm{98}}$ curves), the grown dust in the $\alpha=10^{-4}$ model is mainly concentrated in a broad ring centered at approximately 1~au. A strong pressure maximum in the vicinity of this region, together with dust drift velocities dominated by the gradiental drift (see Appendix~\ref{app:vel}), facilitates the development of a sharp dust ring. 

The total dust-to-gas ratio now strongly deviates from the canonical 1:100 value in most of the gas disk, having strong enhancements up to a factor of 150 in the ring and deep depressions in the rest of the disk, with a mean value averaged over the entire gas disk equal to 6.2.  We note that such high total dust-to-gas ratios in the ring could possibly lead to the development of streaming instability and gravitational collapse of locally overdense regions \citep{2017YangJohansen}. We plan to investigate this scenario in a follow-up study. 
The disk in the $\alpha=10^{-4}$ model is notably colder than the in the higher-$\alpha$ model, although the gas temperature can still reach 1000~K in the sub-au disk regions.
High optical depths in this model cannot offset a much lower viscous heating. 

The maximum size of dust grains in the $\alpha=10^{-4}$ model is much higher than in the other considered models and can reach 100~cm in the dust ring owing to an increased fragmentation radius in the lower-$\alpha$ and colder disk.  The Stokes numbers of grown dust in the disk are substantial (on the order of a few~$\times10^{-1}$ and even reaching 2.0 in the ring), which implies efficient drift of grown dust with respect to gas, as is reflected in the formation of strong dust density enhancements. This effect reduces the total dust-to-gas ratios notably below the initial value of 0.01 in the outer disk regions (an effect also seen in \citet{2014LambrechtsJohansen}).

The pebbles are abundant in this model, and the pebble-to-gas mass ratio can reach 1.45 in the ring. We note that in the ring the values of the pebble-to-gas mass ratio $\zeta_{\mathrm{p2g}}$ are close to those of the total dust-to-gas mass ratio $\zeta_{\mathrm{d2g}}$. This means that most of the dust mass in the ring is found in the form of pebbles. 
Indeed, for the chosen slope of the dust size distribution ($p=3.5$), dust grains near the upper size $a_{\rm max}$ mostly contribute to the total mass budget. In the ring, $a_{\rm max}$ is so large that the mass contributions from the lower-size tail are small for each dust population (see Fig.~\ref{fig:0}). 

Finally, we note the fluctuating behavior of disk temperature and density in the early disk evolution. These fluctuations appear as horizontal variations of the corresponding quantities in Figure~\ref{fig:2}. These features are caused by gravitational instability in the disk, which develops in the early evolution.  Global gravitational perturbations from spiral density waves create alternating radial flows in the disk and cause fluctuations in the mass accretion rate on the star \citep{2016Elbakyan}. Since the gravitational instability is sustained for a longer time in lower-$\alpha$ disks (see Fig.~\ref{fig:1}),  the fluctuations persist longer in lower-$\alpha$ models.

\begin{figure*}
\begin{centering}
\includegraphics[width=2\columnwidth]{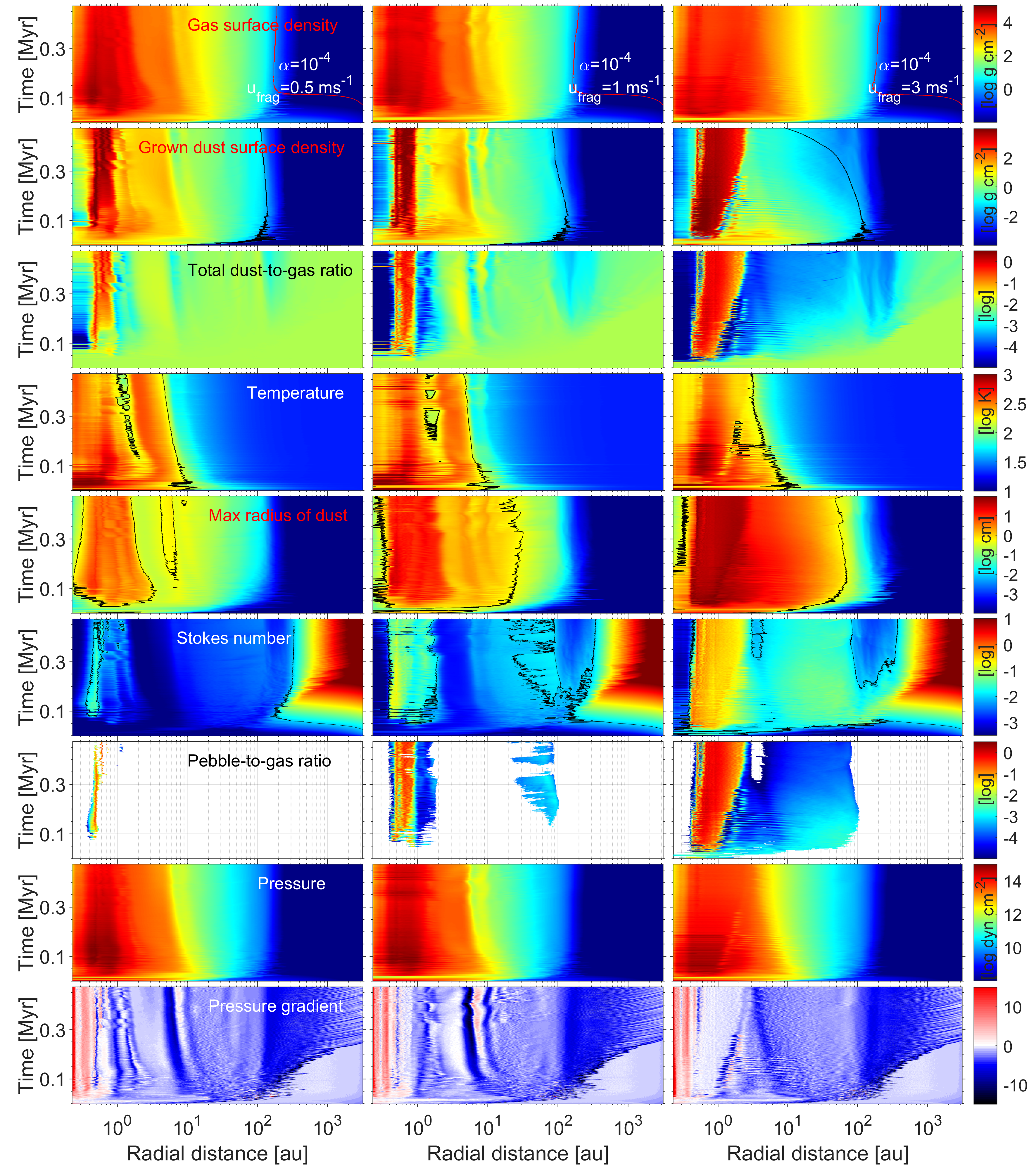}
\par\end{centering}
\caption{\label{fig:8} Similar to Figure~\ref{fig:2}, but for models with $\alpha$=10$^{-4}$ and different values of the dust fragmentation velocity:  $u_{\mathrm{frag}}=0.5$~m~s$^{-1}$ (left column), $u_{\mathrm{frag}}=1$~m~s$^{-1}$ (middle column), and $u_{\mathrm{frag}}=3$~m~s$^{-1}$ (right column). The bottom row shows the pressure gradient, $d \ln {\cal P} / d \ln r$. }
\end{figure*}

\section{Effects of varying dust fragmentation velocity}
\label{sec:param}
In this section, we study how the fragmentation velocity $u_{\rm frag}$ may affect the formation and spatial distribution of pebbles in the $\alpha$=10$^{-4}$~model. We note that $u_{\rm frag}$ enters the definition of the fragmentation barrier in Equation~(\ref{afrag}) and hence influences the maximum size of dust grains in the fragmentation-limited regime of dust growth. 
Many authors have explored either numerically or experimentally the values of fragmentation velocity for different dust particle compositions, sizes, and ice mantles \citep{2008BlumWurm, 2009TeiserWurm, 2009Wada, 2010ZsomOrmel, 2013Wada, 2013Meru, 2014Yamamoto, 2015GundlachBlum, 2017BukhariSyed}. For the silicate grains, the possible values vary between 0.5~m~s$^{-1}$ and 30~m~s$^{-1}$. For our parameter space study, we have chosen two additional values of $u_{\rm frag}$: 0.5~m~s$^{-1}$ and 1.0~m~s$^{-1}$. According to \citet{2019OkuzumiTazaki}, the lowest chosen value of $u_{\rm frag}=0.5$~m~s$^{-1}$ may correspond to bare grains, while the reference value, $u_{\rm frag}=3.0$~m~s$^{-1}$, to dust grains covered with water ice.

\begin{table}
\center
\caption{\label{tab:1}Characteristic parameters for the models}
\resizebox{\columnwidth}{!}{\begin{tabular}{ccccccc}
\hline 
\hline 
Model & $M_{\ast}^{\mathrm{fin}}$ &  $M_{\mathrm{gas}}^{\mathrm{98}}$  & $M_{\mathrm{dust}}^{\mathrm{98}}$ & $M_{\mathrm{peb}}$  & $R_{\mathrm{dust}}^{\mathrm{98}}$  & $R_{\mathrm{gas}}^{\mathrm{98}}$     \tabularnewline
& [$M_{\odot}$]  & [$M_{\odot}$] & [$M_{\oplus}$] & [$M_{\oplus}$] & [AU] & [AU] \tabularnewline
\hline 
$\alpha$=10$^{-2}$ & 0.37 & 0.115 & 260.78 & -      & 287.10 & 1159.9 \tabularnewline
$\alpha$=10$^{-3}$ & 0.32 & 0.172 & 106.47  & 14.24  & 118.87 & 224.73 \tabularnewline
$\alpha$=10$^{-4}$ & 0.27 & 0.225 & 983.22 & 884.60 & 9.30 & 213.99 \tabularnewline
$\alpha$=10$^{-4}$, $u_{\rm frag}=0.5$~m~s$^{-1}$ & 0.30 & 0.192 & 744.26  & 32.17  & 110.45 & 189.32 \tabularnewline
$\alpha$=10$^{-4}$, $u_{\rm frag}=1.0$~m~s$^{-1}$ & 0.31 & 0.187 & 731.49 & 263.64 & 82.32 & 198.82 \tabularnewline
$\alpha$=10$^{-3}$, $\zeta=0.01$ &  0.32 & 0.172 & 573.03 & 75.96  & 118.87 & 224.73 \tabularnewline
$\alpha$=10$^{-4}$, $\zeta=0.01$ &  0.27 & 0.225 & 750.75 & 184.14 & 9.30 & 213.99 \tabularnewline
$\delta=10^{-3}+\alpha_{\rm GI}$ &  0.32 & 0.176 & 492.65 & 23.67  & 116.00 & 230.30 \tabularnewline
$\delta=10^{-4}+\alpha_{\rm GI}$ &  0.29 & 0.205 & 849.85 & 142.57 & 110.45 & 180.27 \tabularnewline
\tabularnewline
\hline 
\end{tabular}}
\center{ \textbf{Notes.} 
$M_{\ast}^{\mathrm{fin}}$ is the final mass of the central star, $R_{\mathrm{gas}}^{\mathrm{98}}$ and $R_{\mathrm{dust}}^{\mathrm{98}}$ are the gas and dust disk radii containing 98\% of total gas and dust mass, respectively, $M_{\mathrm{gas}}^{\mathrm{98}}$ and $M_{\mathrm{dust}}^{\mathrm{98}}$ are the corresponding gas and dust masses. All the values are shown at the end of simulations. $M_{\mathrm{peb}}$} is the total mass of pebbles averaged over the final 100 kyr of simulations. 
\end{table}

In Figure~\ref{fig:8} we compare the azimuthally averaged disk characteristics in the $\alpha=10^{-4}$ model for the three chosen fragmentation velocities: $u_{\mathrm{frag}}$=0.5~m~s$^{-1}$ (left column), $u_{\mathrm{frag}}$=1.0~m~s$^{-3}$ (middle column),  and $u_{\mathrm{frag}}$=3.0~m~s$^{-1}$ (right column). The variations in $u_{\rm frag}$ have a profound effect on the spatial distribution of grown dust and pebbles, as well as on the maximum dust size and Stokes number. The gas surface density, temperature, and pressure are also affected but to a lesser degree and indirectly through the changing opacity. 

The bottleneck effect is naturally present in all three $u_{\rm frag}$ models considered. However, its effect on the radial distribution of grown dust and pebbles is notably distinct.
In the $u_{\rm frag}=3.0$~m~s$^{-1}$ model a monolithic dust ring forms, while the lower-$u_{\rm frag}$ models are characterized by a series of dust rings, with the innermost ring having the highest dust concentration. The origin of multiple ring structures can be seen in the bottom panel of Figure~\ref{fig:8} showing the radial pressure gradient, $d \ln {\cal P} / d \ln r$, which enters the definition of $u_{\rm r,grad}$ in Eq.~(\ref{u_drift}).  Recall that in the $\alpha=10^{-4}$ model $u_{\rm r,grad}$ dominates $u_{\rm adv}$ (see Appendix~\ref{app:vel}), hence the pressure gradient also determines the direction of dust drift in this model. The spatial morphology of $d \ln {\cal P} / d \ln r$ in the models with lower values of $u_{\rm frag}$ is more complex, having several regions where the pressure gradient changes its sign from negative (inward drift) to positive (outward drift). One such region is evident at $\approx$ 10~au. The pressure gradients in the lower-$u_{\rm frag}$ models are also strongly non-monotonic (even when having a similar sign), implying the presence of "traffic jams" that can lead to dust pile-ups.  In the model with a reference value of $u_{\rm frag}=3.0$~m~s$^{-1}$ the spatial morphology of the pressure gradient is smoother, having a clear switch in the sign of $d \ln {\cal P} / d \ln r$ in the sub-au region where the monolithic ring is found. There is another region where the pressure gradient changes sign just outside 1~au and indeed there is a small second ring there, but it disappears after about 0.2 Myr.

The total dust-to-gas ratio $\zeta_{\rm d2g}$ shows strong deviations from the canonical 1:100 value, although the magnitude of this effect decreases with lower values of $u_{\rm frag}$. For instance, the mean dust enhancements (relative to 1:100) in the $u_{\rm frag}=0.5$ and 1.0~m~s$^{-1}$ models are 2.0 and 2.45, while in the $u_{\rm frag}=3.0$~m~s$^{-1}$ model this value can reach 6.2. The overall reduced efficiency of grown dust trapping  in the lower-$u_{\rm frag}$  models can be explained by a smaller maximum dust size and Stokes number of grown dust. For instance,  grown dust in the $u_{\mathrm{frag}}$=0.5~m~s$^{-1}$~model reaches a few centimeters in size, while dust in the $u_{\mathrm{frag}}$=3.0~m~s$^{-1}$~model can grow to 100~cm (see the fifth row in Fig.~\ref{fig:8}). As a result, the maximum $\mathrm{St}$ in the inner disk of the $u_{\mathrm{frag}}$=0.5~m~s$^{-1}$model is close to 0.01, while in the $u_{\mathrm{frag}}$=3.0~m~s$^{-1}$~model the Stokes number exceeds unity in the ring  (sixth row). We note, however, that the dust enhancement for all three values of $u_{\rm frag}$ remains to be quite high in the rings, exceeding the canonical 1:100 value by factors of up to 100-200. There is also a strong spatial stratification in the dust enhancement, with the innermost ring having the highest values.

Pebbles are present for all three values of $u_{\rm frag}$.  However, their spatial distribution is strikingly different. For instance, in the $u_{\rm frag}=0.5$~m~s$^{-1}$~model pebbles are present only in the inner sub-au region. As $u_{\rm frag}$ increases, pebbles begin to occur at larger radial distances, up to $\approx 100$~au for $u_{\rm frag}=3.0$~m~s$^{-1}$. Pebbles are mainly concentrated inside the inner ring structure, where they are trapped as they drift inward.

Finally, we note that if we increase $u_{\rm frag}$ to 10~m~s$^{-1}$ then pebbles start forming in the $\alpha=10^{-2}$ model as well. They are located in an radial annulus with a width of several tens of astronomical units centered at $\approx 40-50$~au.  However, the total mass of pebbles is still the lowest among all considered models and does not exceed 10~$M_\oplus$.

\section{Effects of gravitoturbulence}
\label{sec:vert_turb}
The effect of turbulence induced by the MRI is parameterized in our work in terms of the viscous $\alpha$-parameter. However, gravitational instability is also known to have effects on the dust and gas dynamics that may be similar in some aspects to turbulence \citep{2016Kratter}. High-resolution 3D simulations by \citet{2017RiolsLatter} and \citet{2018RiolsLatter_b} demonstrated that spiral waves in gravitationally unstable disks can stir dust grains in the vertical direction, thus influencing the dynamics and growth of dust. \citet{2020Riols} found that turbulent flows induced by gravitational instability powerfully resist the vertical settling of mm to dm size particles.

The effect of gravitoturbulence on the dynamics of gas and dust in the disk plane is taken into account in FEOSAD self-consistently by calculating the disk self-gravity  (see Eq.~\ref{mom}). We now want to consider the effects that gravitoturbulence  may have on the vertical settling and growth of dust in our models. 
To this end, we introduce the effective $\alpha_{\mathrm{GI}}$ parameter as defined in \citet{2018RiolsLatter}
\begin{equation}
\label{alphaGI}
    \alpha_{\mathrm{GI}} = \frac{G_{r\phi}}{P}\frac{\mathrm{d \ln}r}{\mathrm{d \ln}\Omega},
\end{equation}
where the $(r,\phi)$-component of the gravitational stress tensor has the following form
\begin{equation}
\label{G_rphi}
    G_{r\phi} = \frac{1}{4\pi G}\frac{\partial\Phi}{\partial r}  {1\over r}\frac{\partial\Phi}{\partial \phi}.
\end{equation}
Here, $\Phi$ is the gravitational potential in the disk midplane and the volumetric gas pressure is defined as
\begin{equation}
P = \frac{\Sigma {\cal R} T}{\sqrt{2\pi}H_{\rm g} \mu}
\end{equation}
where $\cal R$ is the universal gas constant and $\mu$ is the mean molecular weight (note that in Eq.~\ref{mom} we used the vertically integrated gas pressure). We have also tried to parameterize the effects of gravitobulence as in \citet{2008Kratter} but found that the above form agrees better with the visual manifestation of gravitational instability in Fig.~\ref{fig:1}. In particular, the Kratter et al. approach predicted notable $\alpha_{\rm GI}$ where the disk was visually almost axisymmetric.

\begin{figure*}
\begin{centering}
\includegraphics[width=2\columnwidth]{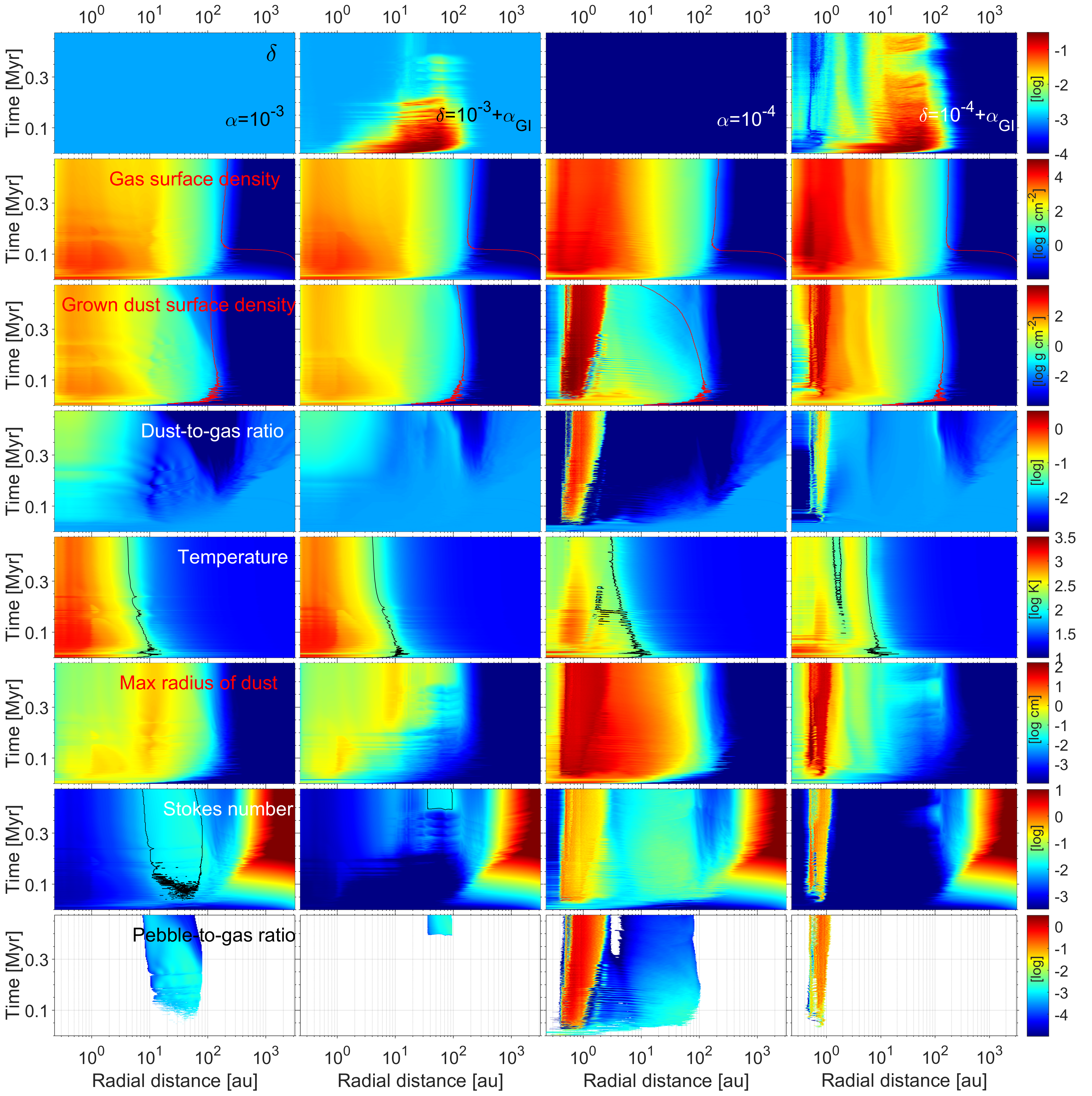}
\par\end{centering}
\caption{\label{fig:12} Similar to Fig.~\ref{fig:2}, but for the $\alpha$=10$^{-3}$~model (left column), $\delta$=10$^{-3}+\alpha_{\mathrm{GI}}$~model (second to the left column), $\alpha$=10$^{-4}$~model (second to the right column), and $\delta$=10$^{-4}+\alpha_{\mathrm{GI}}$~model (right column). In addition, the paramter $\delta = \alpha + \alpha_{\mathrm{GI}}$ is shown in the top row. }
\end{figure*}

We now define the parameter $\delta = \alpha + \alpha_{\mathrm{GI}}$, which is further used instead of $\alpha$ in the definition of the dust vertical scale height $H_{\rm d}$ in Eq.~(\ref{rho:dust}), dust turbulence-induced velocity in Eq.~(\ref{turb_vel}), and dust fragmentation barrier in (Eq.~\ref{afrag}). The latter quantity must be modified because it is derived using the dust turbulent velocity \citep{2016Birnstiel}. Below, only models with a reference value of $u_{\rm frag}=3$~m~s$^{-1}$ are considered. We note that the dust-to-dust collision velocity (Eq.~\ref{turb_vel}) is derived assuming a Kolmogorov-type turbulence. The validity of this expression in the case of gravitoturbulence is discussed in Sect.~\ref{sect:caveats}.

In Figure~\ref{fig:12} we compare the space-time diagrams of the $\alpha$=10$^{-3}$ and $\alpha$=10$^{-4}$~models with those of the $\delta$=10$^{-3}+\alpha_{\mathrm{GI}}$ and $\delta$=10$^{-4}+\alpha_{\mathrm{GI}}$~models. In the top row, we plot the value of $\delta$-parameter for each model, noting that in the fiducial models $\alpha=\delta$ formally. 
Clearly, the inclusion of $\alpha_{\rm GI}$ modifies the effects of turbulence. In particular, gravitoturbulence begins to dominate the magnetorotational turbulence in part of the $\delta=10^{-3}+\alpha_{\rm GI}$ disk and in most of the $\delta=10^{-4}+\alpha_{\rm GI}$ disk. For instance, the mean values of $\alpha_{\rm GI}$ lie in the 0.016--0.026 range for these models, while the maximum values can reach 0.5. For comparison, the viscous $\alpha$-parameter in these models lies in the $10^{-4}-10^{-3}$ range.

The gas disk properties such as the gas surface density (second row) and temperature (fifth row) in the models with and without $\alpha_{\rm GI}$ are weakly affected, while there are notable differences in the dust properties. For instance, $\zeta_{\rm d2g}$ in the $\delta$=10$^{-4}+\alpha_{\mathrm{GI}}$~model notably decreases compared to the $\alpha$=10$^{-4}$~model, having the mean and maximum dust-to-gas mass ratios higher than the 1:100 value by factors 1.6 and 65, respectively. Recall that the corresponding values in the reference $\alpha$=10$^{-4}$~model were 6.2 and  150. The spatial distribution of grown dust is also different. With the inclusion of $\alpha_{\rm GI}$ the spatial concentration of grown dust is less pronounced.  This is the result of a reduced maximum dust size and Stokes number in the disk regions beyond a few astronomical units, causing a slower inward drift of grown dust. 
In the $\delta$=10$^{-4}+\alpha_{\mathrm{GI}}$~model, gravitoturbulence reduces $a_{\rm max}$ and $\mathrm{St}$ everywhere except for the ring, where its effect is much weakened.

Pebbles are now mostly gone in the $\delta$=10$^{-3}+\alpha_{\mathrm{GI}}$ model, except after $t\approx0.4$~Myr when they appear in a narrow region around $50$~au. Gravitational instability diminishes in the outer disk at that late evolution time and $\alpha_{\rm GI}$ drops, allowing dust particles to grow to pebble sizes. In the $\delta$=10$^{-4}+\alpha_{\mathrm{GI}}$ model the spatial distribution of pebbles is also much affected. Now, pebbles are only found in a narrow ring in the sub-au disk region. Overall, the effects of gravitoturbulence on dust stirring and growth have a profound effect on the spatial and temporal distribution of pebbles, making gravitational instability potentially an important effect to account for in models of planetary core growth via pebble accretion.

\section{Surface density, maximum size, and total mass of pebbles}
\label{sect:masses}
\begin{figure}
\begin{centering}
\includegraphics[width=1\columnwidth]{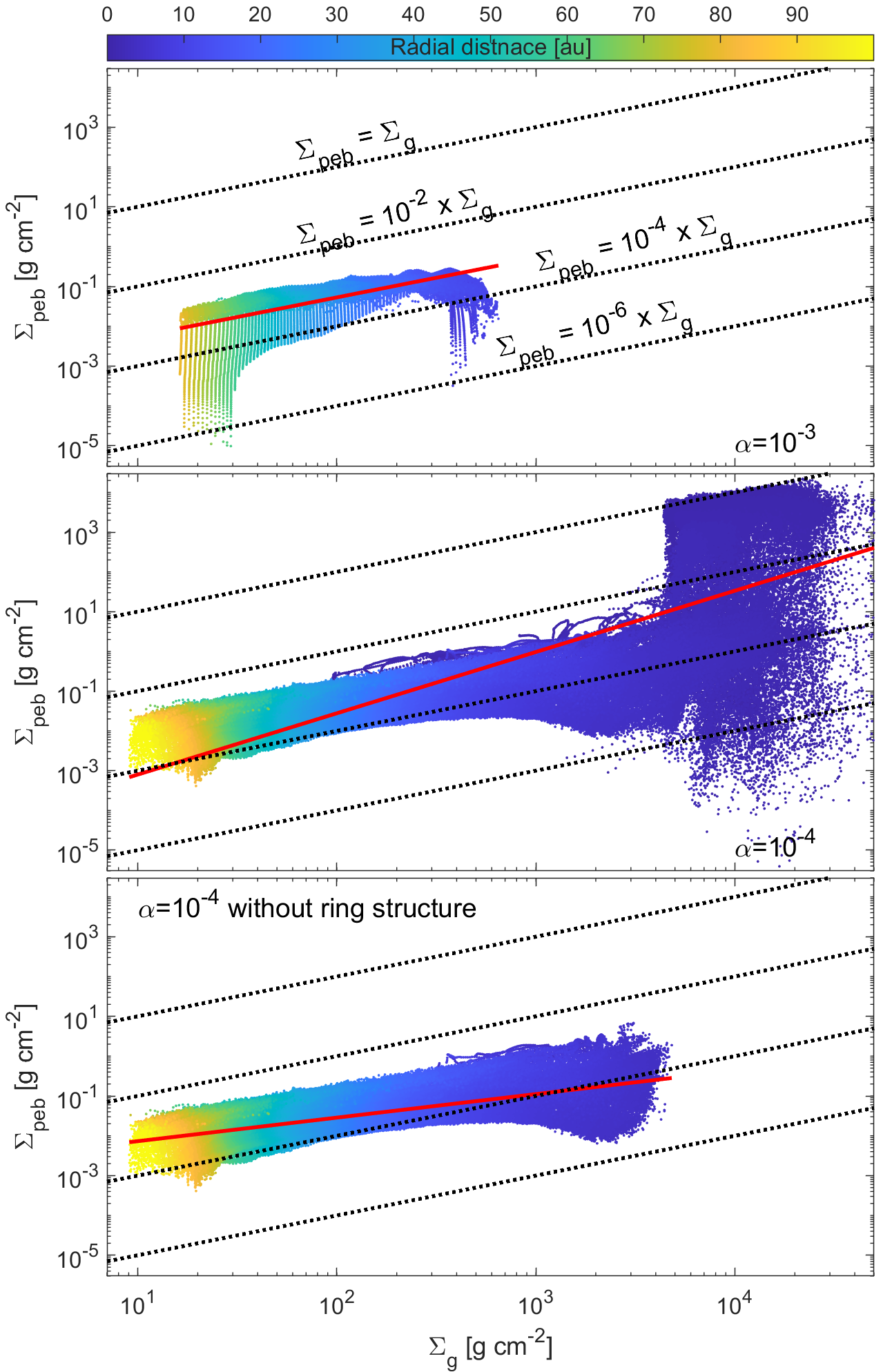}
\par\end{centering}
\caption{\label{fig:8d} The relation between azimuthally averaged pebble ($\overline{\Sigma}_{\rm peb}$) and gas ($\overline{\Sigma}_{\rm g}$) surface densities in the $\alpha=10^{-3}$ model (top panel), $\alpha=10^{-4}$ model (middle panel), and in the $\alpha=10^{-4}$ model but without the ring (bottom panel). The dotted lines show the linear correlations $\overline{\Sigma}_{\rm peb}=C\overline{\Sigma}_{\rm gas}$ with $C$ changing from 1.0 to $10^{-6}$. The red lines show the power-law fitting to the model data.}
\end{figure}

In this section, we analyse the pebbles characteristics in terms of their surface density, maximum size, and total mass. In all cases,  models with $u_{\rm frag}=3.0$~m~s$^{-1}$ and  $\alpha_{\rm GI}=0$ are used, if not stated otherwise. In Figure~\ref{fig:8d} we show the dependence of the azimuthally averaged pebble surface density $\overline{\Sigma}_{\rm peb}$ on the azimuthally averaged gas surface density $\overline{\Sigma}_{\rm g}$. To make the plots, we used the model data from Fig.~\ref{fig:2}  with a time sampling of 5.0~kyr. The top and middle panels correspond to the $\alpha=10^{-3}$ and $\alpha=10^{-4}$~models,  while the bottom panel presents the $\alpha=10^{-4}$~model but excluding the inner ring structure. The plots also carry information about the radial distance in the disk as illustrated by different colors  shown in the color bar. 
The red lines show power-law best fits to the model data. The best-fit coefficients are listed in Table~\ref{tab:2}.

The $\alpha=10^{-3}$~model demonstrates a near-linear correlation between the pebble and gas surface densities, but $\overline{\Sigma}_{\rm peb}$ is factors of $10^{-3}$ to $10^{-4}$ lower than $\overline{\Sigma}_{\rm g}$. Recall that this model showed dust enhancements above the canonical 1:100 value, but this occurred in the inner 10~au of the disk, see Fig.~\ref{fig:2}. However, pebbles in this model are found in a disk annulus between 10 and 80~au so that disk regions with dust concentrations and pebble formation do not match.  Moreover, $\Sigma_{\rm peb}<\Sigma_{\rm d,tot}$ always by definition (see Fig.~\ref{fig:0}). This explains why $\zeta_{\rm p2g}$ in this model is at least a factor of 10 (on average) lower than a reference value of $\zeta_{\rm d2g}=10^{-2}$.

The middle panel in Figure~\ref{fig:8d} considers the $\alpha=10^{-4}$ model.  Unlike the $\alpha=10^{-3}$~case, pebbles in the $\alpha=10^{-4}$ model also exist in the innermost disk, where the dust ring with strong dust concentration is found. This explains why the pebble-to-gas mass ratio $\zeta_{\rm p2g}$  in this model can reach much higher values (up to unity). Because of dust (and pebble) accumulation in the inner disk, the dependence of $\overline{\Sigma}_{\rm peb}$  on $\overline{\Sigma}_{\rm g}$ becomes super-linear. In this context, it is interesting to consider the $\alpha=10^{-4}$ model but with the inner ring structure intentionally excluded from the analysis. The bottom panel in Figure~\ref{fig:8d} presents the corresponding data. The power-law fitting yields a sub-linear function in this case, reinforcing our conclusions above. The $\zeta_{p2g}$ values never exceed 1:100, again because  we now consider disk regions where dust enhancements are absent, see Fig.~\ref{fig:2}.

\begin{table}
\center
\caption{\label{tab:2}Coefficients for the best-fit power-law curves in Figure~\ref{fig:8d}.}
\begin{tabular}{ccc}
\hline 
\hline 
Model & \multicolumn{2}{c}{$10^{p_2}r^{p_1}$}  \tabularnewline
& $p_1$ &  $p_2$  \tabularnewline
\hline 
$\alpha$=10$^{-3}$          &0.99  &-3.25  \tabularnewline
$\alpha$=10$^{-4}$          &1.54  &-4.63  \tabularnewline
$\alpha$=10$^{-4}$ no ring  &0.59  &-2.73  \tabularnewline
\hline 
\end{tabular}
\end{table}

\begin{figure}
\begin{centering}
\includegraphics[width=1\columnwidth]{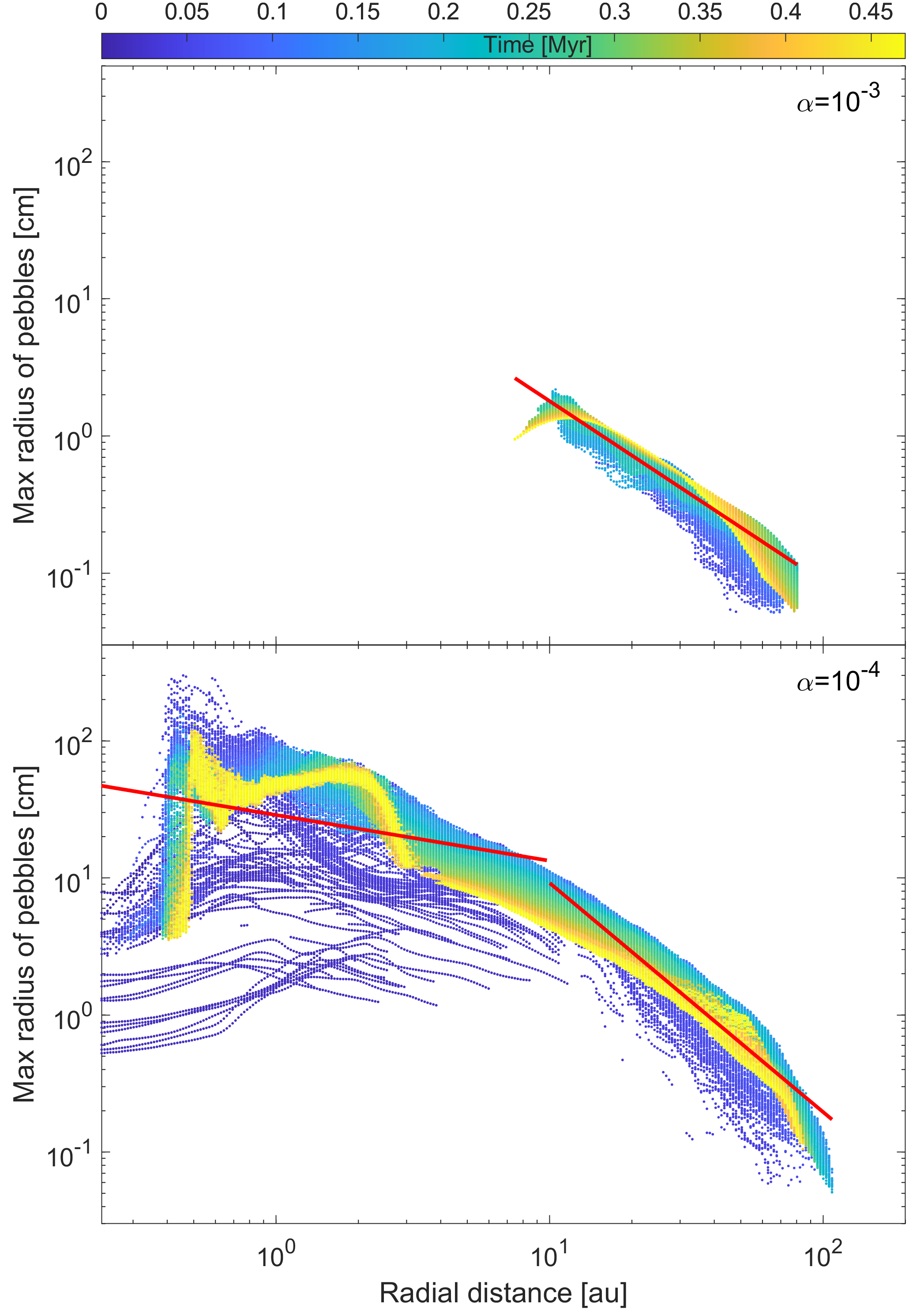}
\par\end{centering}
\caption{\label{fig:8dd} The radial distribution of azimuthally averaged maximum pebble size in the $\alpha=10^{-3}$ (top panel) and $\alpha=10^{-4}$ (bottom panel) models. The age of the system is shown with in the color bar. The red lines show the power-law fitting to the model data.}
\end{figure}

The radial dependence of the  azimuthally averaged maximum pebble size is shown in Figure~\ref{fig:8dd} for models with $\alpha=10^{-3}$ and $10^{-4}$. Different colors provide information on the age of the disk, as shown in the color bar. The red lines show the power-law fits to the model data. The polynomial coefficients for the best-fit curves are listed in Table~\ref{tab:3}. Pebbles in the $\alpha$=10$^{-3}$ model are only present at distances beyond 10~au and reach a maximum size of a few centimeters. The maximum size declines with radius faster than $1/r$, which likely reflects the efficient inward drift of grown dust.  In the $\alpha$=10$^{-4}$ model, owing to the formation of a ring structure in the inner disk, pebbles are present almost in the entire disk. They can now reach a meter in size in the sub-au disk regions, meaning that they can technically be considered as boulders. The slope of the pebble size distribution as a function of radial distance cannot be represented by a single power-law function and is much shallower in the inner 10~au than at larger distances. This can be explained by the saturation of dust growth when pebbles attain its maximum size set by the fragmentation barrier in the inner disk regions.

\begin{table}
\center
\caption{\label{tab:3}Coefficients for the best-fit power-law curves in Figure~\ref{fig:8dd}.}
\begin{tabular}{ccc}
\hline 
\hline 
Model & \multicolumn{2}{c}{$10^{p_2}r^{p_1}$}   \tabularnewline
& $p_1$ &  $p_2$  \tabularnewline
\hline 
$\alpha$=10$^{-3}$          &-1.32  &1.57 \tabularnewline
$\alpha$=10$^{-4}$, $r<10$~au          &-0.33  &1.46 \tabularnewline
$\alpha$=10$^{-4}$, $r>10$~au          &-1.67  &2.63 \tabularnewline
\hline 
\end{tabular}
\end{table}

\begin{figure}
\begin{centering}
\includegraphics[width=1\columnwidth]{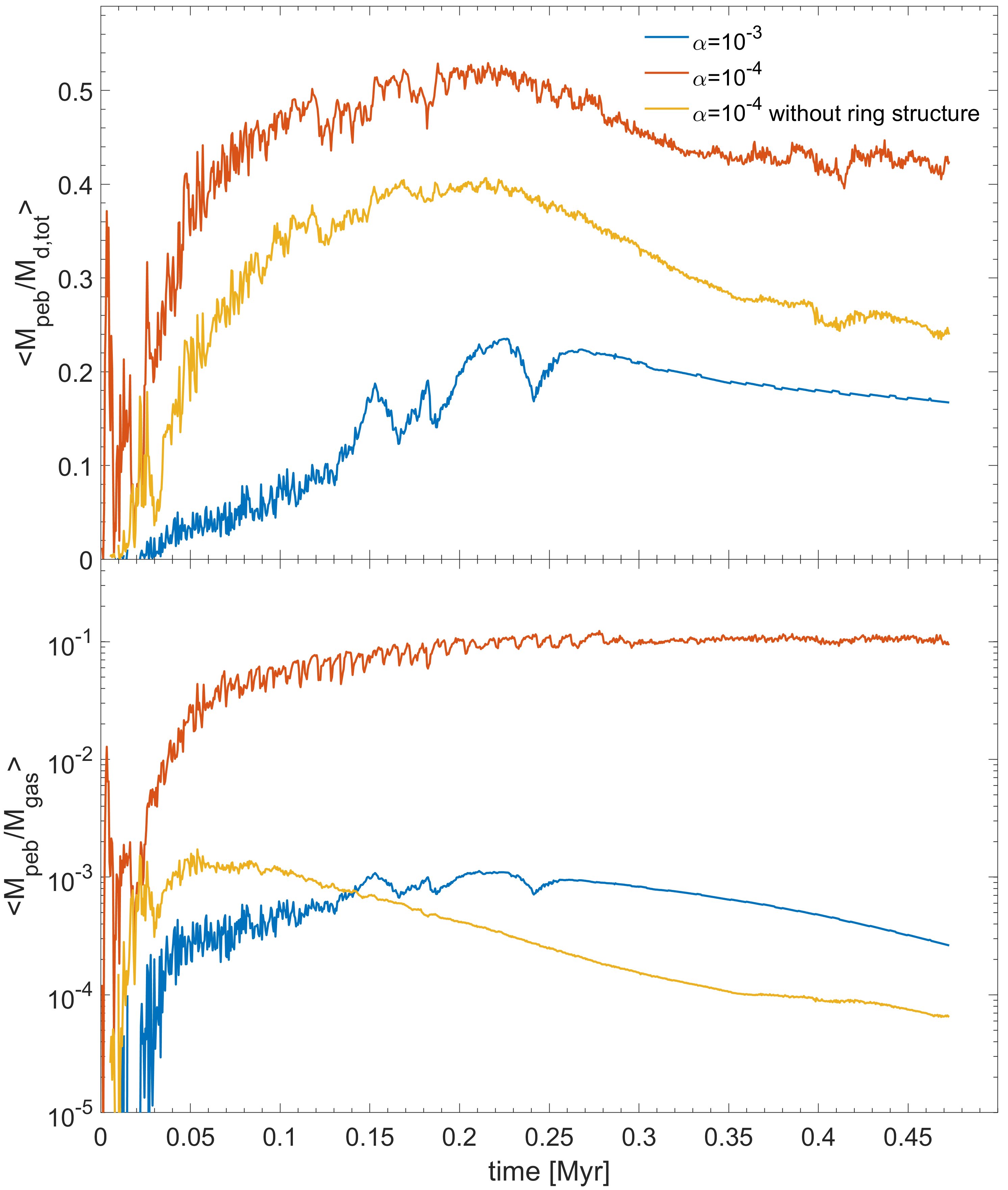}
\par\end{centering}
\caption{\label{fig:8h} Time evolution of the pebble-to-total dust mass ratio (top panel), and pebble-to-gas mass ratio (bottom panel), both averaged over the disk regions where pebbles are found. }
\end{figure}

It is interesting to consider the time evolution of the pebble-to-dust and pebble-to-gas mass ratios in the entire disk. In the top panel of Figure~\ref{fig:8h}, we show the time evolution of $\langle M_{\rm peb}/M_{\rm d,tot} \rangle$ in our models, where $M_{\rm peb}$ and $M_{\rm d,tot}$ are the masses of pebbles and dust (both small and grown), respectively, and the brackets denote averaging over the disk regions where pebbles are found. We note that we first find the ratio in each computational cell where pebbles are present and then perform the averaging, and not vice versa.
This ratio can be regarded as the mass fraction of pebbles in the total mass of dust. 

During the early phase ($t\lessapprox0.2$~Myr), $\langle M_{\rm peb}/M_{\rm d,tot} \rangle$ increases as a result of efficient dust growth \citep[see also][]{2018VorobyovAkimkin}. At around 0.2~Myr almost half of the total dust mass is converted to pebbles in the pebble-forming regions of the $\alpha=10^{-4}$ model.
In the $\alpha=10^{-3}$ model this fraction is lower because of a lower fragmentation barrier $a_{\rm frag}$, which limits dust growth and conversion of dust to pebbles. Later, however, the fraction of pebbles in the total dust mass begins to decline in both models. This effect reflects the overall change in the local disk conditions when $a_{\rm frag}$ declines with time, causing the mass of pebbles to decline as well in the fragmentation-limited regime of dust growth.

In the bottom panel of Figure~\ref{fig:8h}, we show the time evolution of $\langle M_{\rm peb}/M_{\rm gas} \rangle$, where $M_{\rm gas}$ is the mass of gas  and the averaging is done over the disk regions where pebbles are found. In the $\alpha=10^{-3}$~model, the averaged pebble-to-gas mass ratio reaches $\approx10^{-3}$ at 0.05~Myr and slowly decreases afterward. In contrary, $\langle M_{\rm peb}/M_{\rm gas} \rangle$ in the $\alpha=10^{-4}$~model, continues to grow and saturates at $\approx10^{-1}$ after $t=0.2$~Myr. The difference can be explained by a strong pressure maximum in the inner disk of the $\alpha=10^{-4}$ model, which efficiently traps pebbles  as they drift inward.  In the $\alpha=10^{-3}$ model, pebbles are only found beyond 10~au; recall that $a_{\rm max}$ and/or $\mathrm{St}$ drops there below the values appropriate for pebbles. Their inward drift and destruction in the inner regions cases the  $\langle M_{\rm peb}/M_{\rm gas} \rangle$ ratio to decline with time.
Interestingly, if the ring structure is excluded from the
analysis of the $\alpha=10^{-4}$ model, both  $\langle M_{\rm peb}/M_{\rm d,tot} \rangle$ and  $\langle M_{\rm peb}/M_{\rm gas} \rangle$ decrease substantially and start declining with time as in the $\alpha=10^{-3}$ model, thus proving a dominant contribution of the ring structure to the pebble mass budget.

\begin{figure}
\begin{centering}
\includegraphics[width=1\columnwidth]{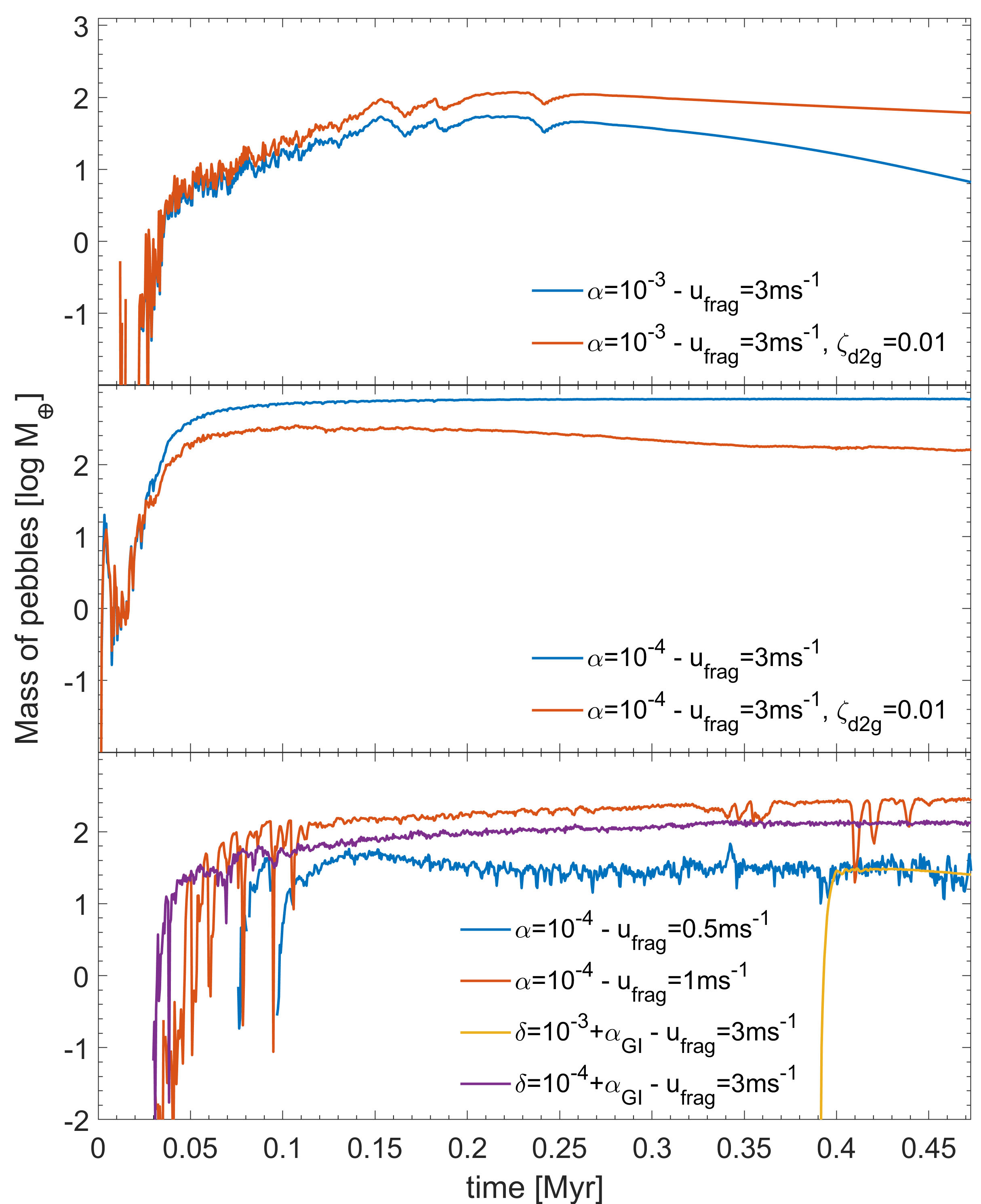}
\par\end{centering}
\caption{\label{fig:8b} Total mass of pebbles in the disk as a function of time for all model realizations considered in this work. Top and middle panels consider the impact of a fixed dust-to-gas mass ratio on the estimates of pebble mass, while the bottom panel shows the results of the parameter space study.}
\end{figure}

Finally, we present in Figure~\ref{fig:8b} the total mass of pebbles $M_{\rm peb}$ in the disk as a function of time in all models, for which pebbles were found. In addition, we consider two models in which the pebble mass was derived not from the actual dust dynamics equations but simply assuming for each grid cell a conversion ratio of 1:100 between dust and gas masses. This is done to illustrate the extent of an error in the pebble mass estimates that may occur when using the standard dust-to-gas mass conversion. These models are denoted as $\zeta_{\rm d2g} = 0.01$ in the Figure. The top and middle panels compare the pebble masses in the $\alpha=10^{-3}$ and $10^{-4}$ models with a reference value of $u_{\rm frag}$=3~m~s$^{-1}$ against the models where the pebble mass was calculated using the 1:100 conversion factor.
Clearly, the use of the simple conversion method introduces a notable error in the total pebble mass estimates. In the $\alpha=10^{-4}$ model, the 1:100 conversion underestimates $M_{\rm peb}$ because pebbles are effectively trapped in the ring, leading to their accumulation in the innermost disk regions, an effect that cannot be reproduced by the simple 1:100 conversion factor. The $\alpha=10^{-3}$ model shows the opposite effect when the 1:100 conversion overestimates the total pebble mass. This occurs owing to efficient pebble drift to the inner disk regions, where the conditions for pebbles are not favourable, and they are transformed into smaller dust particles via collisions.    

The bottom panel present the total pebble mass for models with varying $u_{\rm frag}$ and also in models where the effects of gravitoturbulence on dust growth were taken into account. It is interesting to compare the final pebble masses at the end of numerical simulations at 0.5~Myr in all models considered. These values are provided in the fifth column of Table~\ref{tab:1}. Since $M_{\rm peb}$ sometimes shows short-term variations, we averaged $M_{\rm peb}$ over the last 100~kyr of computed evolution. The highest and lowest masses of pebbles were found in the $\alpha=10^{-4}$ and $10^{-3}$ models with $v_{\rm frag}=3$~m~s$^{-1}$, respectively. The effect of gravitoturbulence significantly reduces the pebble mass in the $\alpha=10^{-4}$ model. The same is true when the dust fragmentation velocity is reduced.

The efficiency of protoplanet growth via accretion of pebbles depends on the pebble mass flux past a protoplanetary seed \citep[e.g.,][]{2019LambrechtsMorbidelli}. We do not calculate pebble fluxes in this work because of the complexity of the flux pattern in the low-$\alpha$ models, leaving this study for a follow-up paper,  but it is still worth estimating if the total pebble mass is sufficient for the formation of planets of different type in the models considered. If we take a 10\% accretion efficiency of pebbles by a protoplanetary seed \citep{2018OrmelLiu}, then
perhaps only the $\alpha=10^{-4}$ models (regardless of $u_{\rm frag}$ and $\alpha_{\rm GI}$) can form Earth-type protoplanets, considering that pebbles are spread over a large disk area in the $\alpha=10^{-3}$ models. For the lowest value of $u_{\rm frag}=0.5$~m~s$^{-1}$ the formation of giant planets via the core collapse scenario becomes unlikely because solid cores have to be sufficiently massive to accrete gaseous atmospheres of giant planets ($\sim 10~M_\oplus$). Gravitoturbulence makes giant planets more difficult to form but the total pebble mass ($130~M_\oplus$) may still be sufficient to form them in the $\alpha=10^{-4}$ model, considering that pebbles are concentrated in a narrow ring, see Fig.~\ref{fig:12}.
Thus, our numerical experiments demonstrate the crucial importance of $u_{\rm frag}$ and, hence, of dust composition and ice coating on the efficiency of planet formation.

\section{Model caveats}
\label{sect:caveats}

{\it Variations in disk masses}. In this study, we considered only one disk evolution model. Variations in the initial mass and angular momentum of pre-stellar cores would result in different disk masses \citep{2011Vorobyov}, which can affect dust growth and pebble formation efficiency. In this work, the disk is rather massive, $\approx 0.1-0.2~M_\odot$, with a consequence that dust can grow to a larger size owing to an increase in $a_{\rm frag}$ (see Eq.~\ref{afrag}). In lower masses disks pebble formation may not be that efficient. A disk parameter study should be performed in the future to better understand the pebble formation efficiency in disks with different masses.

{\it Dust fragmentation velocity}. In this study, we adopted $v_{\rm frag}$ to be a constant of time and space. This is likely an oversimplification as the dust fragmentation properties are known to depend on the presence/absence of ice mantles \citep{2019OkuzumiTazaki}. As a result, $v_{\rm frag}$ is expected to be lower inside the water snowline, which can affect the pebble formation efficiency in the dust rings around 1~au as was shown in Sect.~\ref{sec:param}. A future study should include radial (and azimuthal) variations in $v_{\rm frag}$ depending on the composition of ice mantles as was done in, e.g., \citep{2021Molyarova}. 

{\it Disk magnetic winds}. Our disk evolution model includes two mechanisms of mass and angular momentum transport: turbulent viscosity and gravitational instability, but neglects the effects of disk winds, which can be important in the context of disk and dust evolution \citep[e.g.,][]{2021Taki}. This effect should be considered in the future.

{\it Longevity of the dust rings}. Our estimates show that conditions in the dust rings around 1~au are favorable for the development of the streaming instability \citep{2017YangJohansen}, which implies that  the pebbles that accumulate in these structures will be converted to planetesimals over several tens to hundreds of orbital periods. This process would significantly change the form and dust content in the inner ring, a process that we plan to investigate in a follow-up paper. The first generation of planetesimals may also seed a first protoplanet in an inside-out formation scenario advocated by \citet{2014ChatterjeeTan}.   Nevertheless, this does not invalidate our main finding that large amounts of pebbles can accumulate in the inner regions of low-$\alpha$ disks. However, the time that the pebbles would spend in the ring would be shorter because of efficient process of pebble to planetesimal conversion. 

{\it Accretion bursts}. The $\alpha$-parameter in our models is a constant of time and space. In the low-$\alpha$ models,  however, a thermal or magnetorotational instability may be triggered if gas temperature (and ionization) exceeds a certain threshold in the ring \citep{2020VorobyovKhaibrakhmanov}. This may affect the pebble accumulation efficiency in the ring, dumping part of the accumulated gas and dust on to the star, a process investigated in more detail in \citet{2022Kadam}.

{\it Gravitoturbulence}. 
As was found in \citet{2010Vorobyov_alpha}, the  $\alpha$-parameterization of the effects of disk self-gravity is valid for as long as the disk mass does not exceed 20\%--30\% that of the star. However, the GI-induced turbulence may have spectrum that is different from a Kolmogorov-type turbulence, with the assumption of which the dust-to-dust collision velocity $v_{\rm turb}$ was derived (see Eq.~\ref{turb_vel}). For instance, \citet{2021BaehrZhu} found that the GI-induced turbulence may be slightly anisotropic. The closed-box simulations of \citet{2017RiolsLatter} indicate that the spectrum of gravitoturbulence becomes steeper than that of Kolmogorov’s for wavenumbers smaller than $ 10/H_{\rm g}$. Less turbulent energy at shorter wavelengths implies that gravitoturbulence may be less efficient in sustaining random dust-to-dust collisions. However, \citet{2019BoothClarke} argued that the deviation from the Kolmogorov spectrum may be caused by numerical dissipation. Overall, we may indeed overestimate the dust-to-dust turbulent velocity owing to gravitoturbulence when using the expression derived for the Kolmogorov spectrum, but the magnitude of this effect is presently difficult to constrain.

\section{Conclusions}
\label{sec:concl}
We studied the evolution of self-gravitating protoplanetary disks with different but spatially and temporally constant values of the viscous $\alpha$-parameter, starting our simulations from the gravitational collapse of a prestellar core and ending after about half a million years of disk evolution. We used the hydrodynamics code FEOSAD, which employs the thin-disk approximation and features the co-evolution of gas and dust, including dust growth and  back reaction of dust on gas. We cover both the Epstein and Stokes regimes of dust dynamics using the Henderson drug coefficient.  The initial dust-to-gas mass ratio $\zeta_{\rm d2g}$ in the collapsing cloud core was set equal to the canonical 1:100 value and all dust was initially in the form of sub-$\mu m$ grains.  The impact of turbulent viscosity, dust fragmentation velocity $u_{\rm frag}$, and gravitoturbulence owing to disk gravitational instability (parameterized in terms of the effective parameter $\alpha_{\rm GI}$) on the abundance and spatial distribution of pebbles in the disk is studied in detail. Our main results can be summarized as follows.

-- Turbulent viscosity as parameterized by the $\alpha$-parameter strongly affects the disk evolution. In the $\alpha=10^{-2}$ model, the viscous $\alpha$-value is greater than the $\alpha_{\rm GI}$-value  in most of the disk, making this model viscosity-dominated. The gas and dust spatial distributions are rather similar, with the dust-to-gas mass ratio deviating from the 1:100 value by no more than 30\%.  Stokes numbers do not exceed a few $\times 10^{-3}$, dust drift is controlled by radial gas advection and not by pressure gradients, which explains the lack of dust accumulation in the disk. The maximum size of dust grains $a_{\rm max}$ does not exceed 0.5~mm and pebbles are completely absent in this model.

-- The evolution of gas and dust in the $\alpha=10^{-3}$ and $10^{-4}$ models is different. Gravitational instability begins to dominate in these models, in particular for $\alpha=10^{-4}$, so that $\alpha_{\rm GI}$ becomes greater than the viscous $\alpha$-parameter in most of the disk and throughout the considered evolution period. A bottle neck effect caused by radially varying gravitational torques in a gravitationally unstable disk makes gas and dust accumulate in the innermost disk regions. Stokes numbers now exceed 0.1 and the dust-to-gas mass ratio can be greater than the 1:100 value by more than a factor of 100. Pebbles are abundant, but in the $\alpha=10^{-4}$ model they are mostly concentrated in a ring around 1~au.

-- The abundance and spatial distribution of pebbles is found to be sensitive to the adopted value of the dust fragmentation velocity. A decrease of $u_{\rm frag}$ from 3.0~m~s$^{-1}$ to 0.5~m~s$^{-1}$ makes pebbles disappear throughout most of the disk radial extent, except for the inner sub-au region where they survive in a narrow ring.
This result demonstrates the importance of coupling of dust growth models with the dynamics and phase transitions of volatiles, as was done in, e.g., \citet{2021Molyarova}.

-- Gravitoturbulence can also have a profound effect on pebbles, reducing their total mass and shrinking the spatial extent where they can be found. This can be explained by an increase in the turbulent velocity of dust grains when gravitoturbulence is taken into account, thus effectively reducing the fragmentation barrier and the ability for dust grains to grow to pebble sizes.

-- The integrated mass of pebbles in the disk is nevertheless sufficient to form Earth-type planets for the considered parameter space and also giant planets, if $u_{\rm frag}\ge 1.0$~m~s$^{-1}$ and $\alpha<10^{-3}$. We provide the fitting functions to the surface density and maximum size of pebbles, which may be useful in the planet formation theories. We also show that using the standard 1:100 dust-to-gas mass conversion criterion can lead to substantial errors in the pebble mass estimates.

\begin{acknowledgements}
We are thankful to the anonymous referee for useful comments that helped to improve the manuscript.
E.I.V., V.E., and A.M.S. acknowledge support by the Ministry of Science and Higher Education of the Russian Federation (State assignment in the field of scientific activity, Southern Federal University, VnGr /2020-03-IF, 2020).
S.O.P. acknowledges support from the Ministry of Higher Education and Science of Russian Federation, project of LIH SB RAS FWGG-2021-0001
V.G.E. acknowledges the Swedish Institute for a visitor grant allowing to conduct research at Lund University. AJ was supported by the Swedish Research Council (grant 2018-04867), the Knut and Alice Wallenberg Foundation (grant 2012.0150) and the European Research Council (ERC Consolidator Grant 724687-PLANETESYS). ML was supported by the Knut and Alice Wallenberg Foundation (grant 2012.0150).  The simulations were performed on the Vienna Scientific Cluster (VSC-4).
\end{acknowledgements}

\bibliographystyle{aa}
\bibliography{ref_base}

\begin{appendix}

\section{Drift velocity of dust} 
\label{app:vel}
\begin{figure*}
\begin{centering}
\includegraphics[width=2\columnwidth]{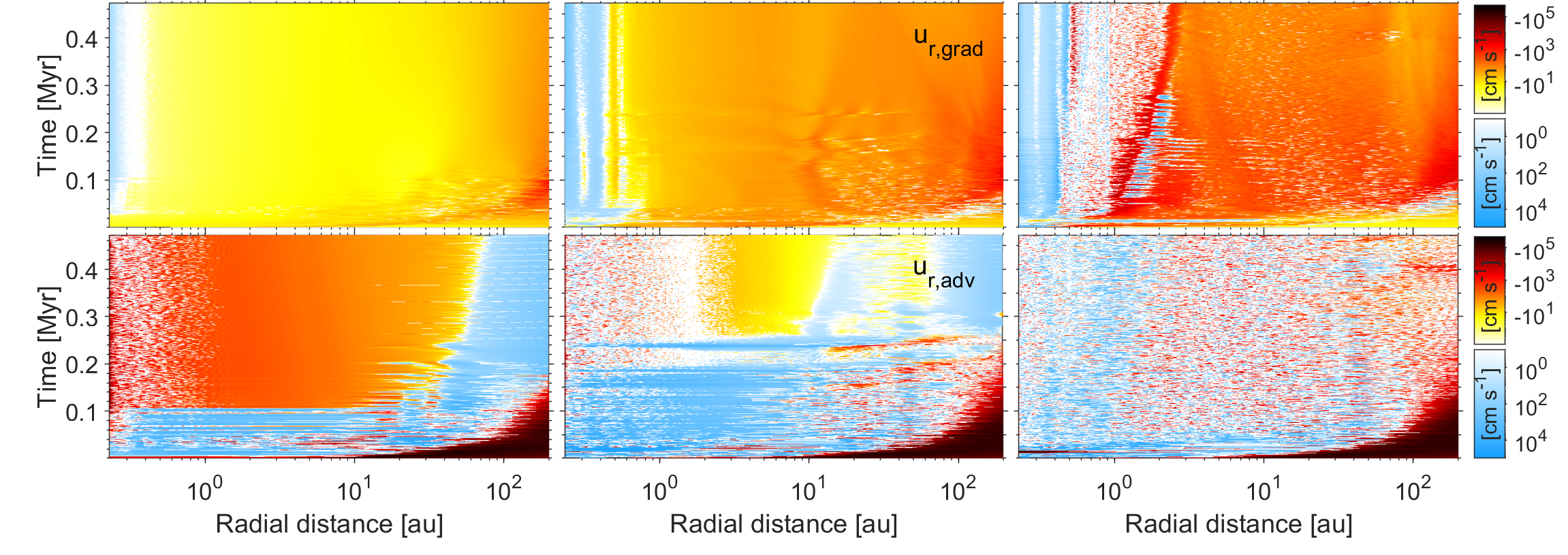}
\par\end{centering}
\caption{\label{fig:13} Space-time diagrams of the azimuthally averaged components of the grown dust drift velocity $u_{r,\mathrm{grad}}$ (top row) and  $u_{r,\mathrm{adv}}$ (bottom row) for three models with $\alpha=10^{-2}$ (left column), $\alpha=10^{-3}$ (middle column) and  $\alpha=10^{-4}$ (right column).
The hot colors corresponds to the inward drift, while the cold colors to the outward drift.
}
\end{figure*}

The radial velocity of dust in a protoplanetary disk can be decomposed into the gradiental and advective 
components,  $u_{\rm r,\mathrm{grad}}$ and $u_{\rm r,\mathrm{adv}}$, as expressed by Eqs.~(\ref{u_drift}) and (\ref{u_adv}), respectively. We calculate these velocities based on the model's known data, average them over the azimuthal extent of the disk, and plot them in Figure~\ref{fig:13} as a function of time and radial distance. Clearly, $u_{\rm r,\mathrm{adv}}$ dominates in the $\alpha=10^{-2}$ model. Since this model is characterized by rather low $\mathrm{St} < 0.01$, grown dust is basically advected along with gas and deviations of $\zeta_{\rm d2g}$ from the canonical value of 1:100 are small.  

On the other hand, $u_{r,\mathrm{grad}}$ dominates the dust drift velocity in the $\alpha=10^{-3}$ and $10^{-4}$ models. Together with substantial $\mathrm{St}\gg 0.01$ this implies decoupling of grown dust dynamics from that of gas, followed by accumulation of dust in pressure maxima. The change of sign in $u_{r,\mathrm{grad}}$ in the innermost disk regions reflects a converging flow of grown dust towards the pressure maximum.

\section{Semi-analytical explanation of the bottle neck effect}
\label{bottle:neck}

We illustrate the formation of a gas ring in our low-$\alpha$ models using a simplified one-dimensional approach. The evolution of an axisymmetric, geometrically thin, viscous gaseous disk with surface density $\Sigma_{\rm g}(r,t)$ and angular velocity $\Omega(r)$ can be described by the following viscous equation \citep{1981Pringle}
\begin{equation}
\label{app_dSigma}
    \frac{\partial \Sigma_{\rm g}}{\partial t} + \frac{1}{r} \frac{\partial}{\partial r} \left[ \frac{1}{(r^2\Omega)'} \frac{\partial}{\partial r} (\nu\Sigma_{\rm g} r^3 \Omega')\right] = 0,
\end{equation}
where the primes stand for the differentiation with respect to radius and $\nu$ is the kinematic viscosity. Making use of the continuity equation for $\Sigma_{\rm g}$ and noting that the mass transport rate through an annulus with radius $r$ is $\dot{M}=-2\pi r\Sigma v_{\mathrm{r}}$, where $v_{\rm r}$ is the gas radial velocity, we can write
\begin{equation}
\label{app_dSigma_mdot}
    \frac{\partial\Sigma_{\rm g}}{\partial t} = {1 \over r} \frac{\partial}{\partial r} \bigg( \frac{\dot{M}}{2\pi} \bigg),
\end{equation}
where the mass transport rate can now be expressed as
\begin{equation}
\label{app_mdot}
     \dot{M} = -\frac{2\pi}{(r^2\Omega)'} \frac{\partial}{\partial r} (\nu\Sigma_{\rm g} r^3 \Omega').
\end{equation}

Using the Shakura-Sunyaev $\alpha$-prescription, the kinematic viscosity can be expressed as
\begin{equation}
\nu = \alpha_{\mathrm{eff}} c_{\mathrm{s}} H_{\rm g}.
\end{equation}
If we only considered viscous evolution, then $\alpha_{\rm eff}=\alpha$ and in our models $\alpha_{\rm eff}$ would be a constant of time and space (either $10^{-2}$ or $10^{-3}$ or $10^{-4}$). However, our models include disk self-gravity as well, the effect of which can also be parameterized in terms of an effective viscosity for moderate disk-to-star mass ratios \citep{2010Vorobyov_visc}.
Therefore, the effective  parameter $\alpha_{\mathrm{eff}}$ should take into account the effects of turbulent viscosity and gravitational instability and be expressed in the following form
\begin{equation}
\alpha_{\mathrm{eff}}=\alpha + \alpha_{\mathrm{GI}},
\label{alpha_eff}
\end{equation}
where $\alpha$ is a usual (and constant in our models) parameter to represent the strength of turbulent viscosity in the disk and $\alpha_{\mathrm{GI}}$ is a viscous parameter that describes the mass and angular momentum transport through gravitational torques in the disk. The latter quantity is defined by Equation~(\ref{alphaGI}). We note that $\alpha_{\rm eff}$ is similar to $\delta$ introduced in Sect.~\ref{sec:vert_turb}. The only difference is that $\delta$ is used in the dust growth equations, while $\alpha_{\rm eff}$ is used here to solve Eq.~(\ref{app_dSigma_mdot}).

\begin{figure}
\begin{centering}
\includegraphics[width=1\columnwidth]{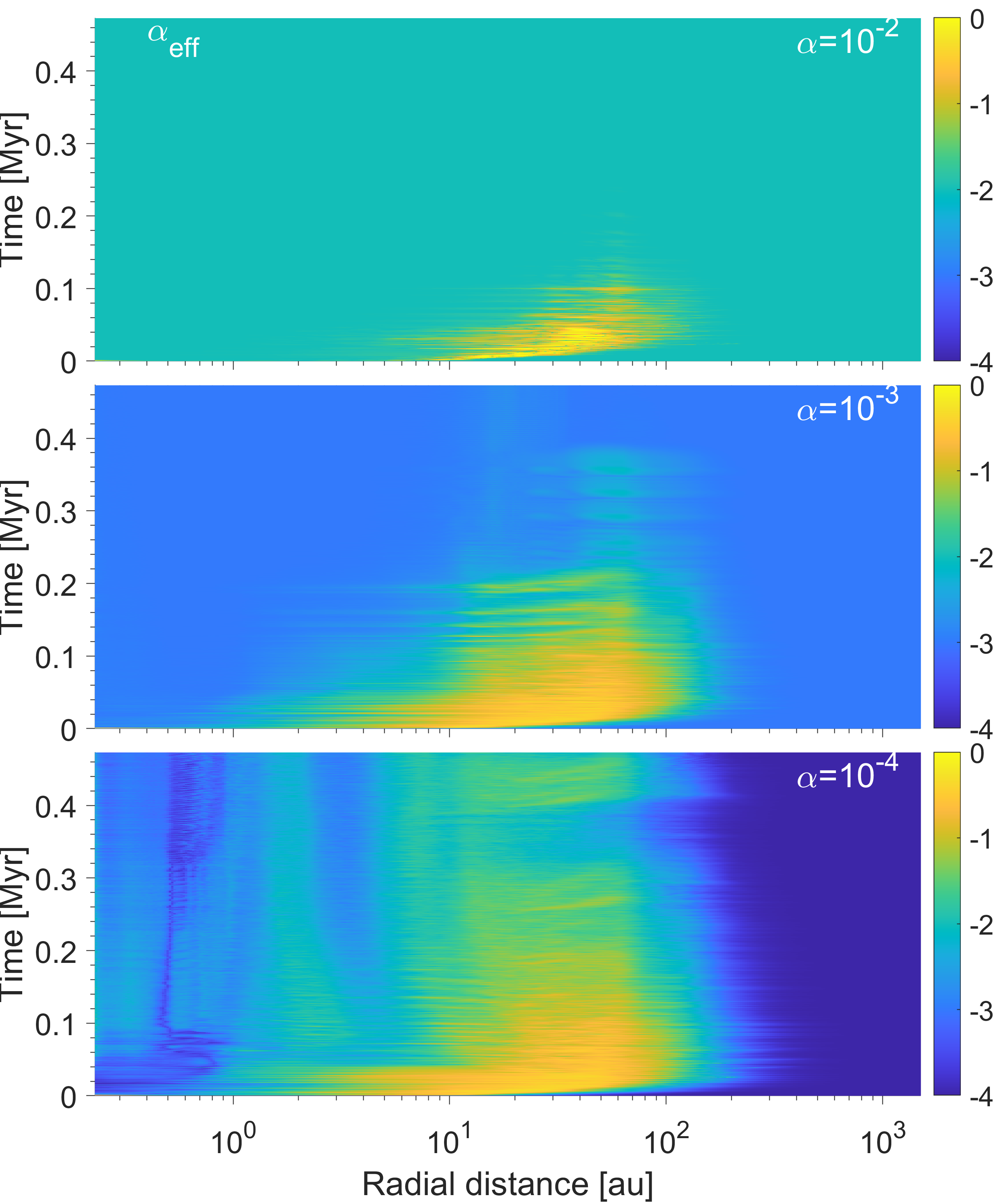}
\par\end{centering}
\caption{\label{fig:13a} Space-time diagram of azimuthally averaged $\alpha_{\rm eff}$ in our models. The top, middle, and bottom panels correspond to $\alpha=10^{-2}$, $10^{-3}$ and $10^{-4}$ models, respectively.}
\end{figure}

Figure~\ref{fig:13a} presents the azimuthally averaged space-time diagrams of $\alpha_{\rm eff}$ in the three models with constant values of the viscous $\alpha$-parameter. In the $\alpha=10^{-2}$ model, the effective parameter $\alpha_{\rm eff}$ is constant almost everywhere except for the initial 0.1~Myr, during which a notable contribution from disk gravitational instability can be seen between 10 and 100~au. This implies that viscous torques are dominant through most of the considered evolution. In the $\alpha=10^{-3}$ model, the contribution from gravitational instability extends to 0.25~Myr and also to smaller distances. In contrast, disk gravitational instability is almost entirely dominant in  the $\alpha=10^{-4}$ model. Furthermore, $\alpha_{\rm eff}$ is highly nonhomogeneous, having the highest values in the intermediate and outer disk regions, where gravitational instability is strongest (see Fig.~\ref{fig:1}), and lowest values in the inner disk regions, where gravitational instability is suppressed owing to high temperature and strong shear. These radial variations of $\alpha_{\rm eff}$ represent the essence of the bottle neck effect in the low-$\alpha$ models -- the material is transported from the outer to the inner disk regions with a decreasing efficiency, leading to its accumulation at around 1~au.

\begin{figure}
\begin{centering}
\includegraphics[width=1\columnwidth]{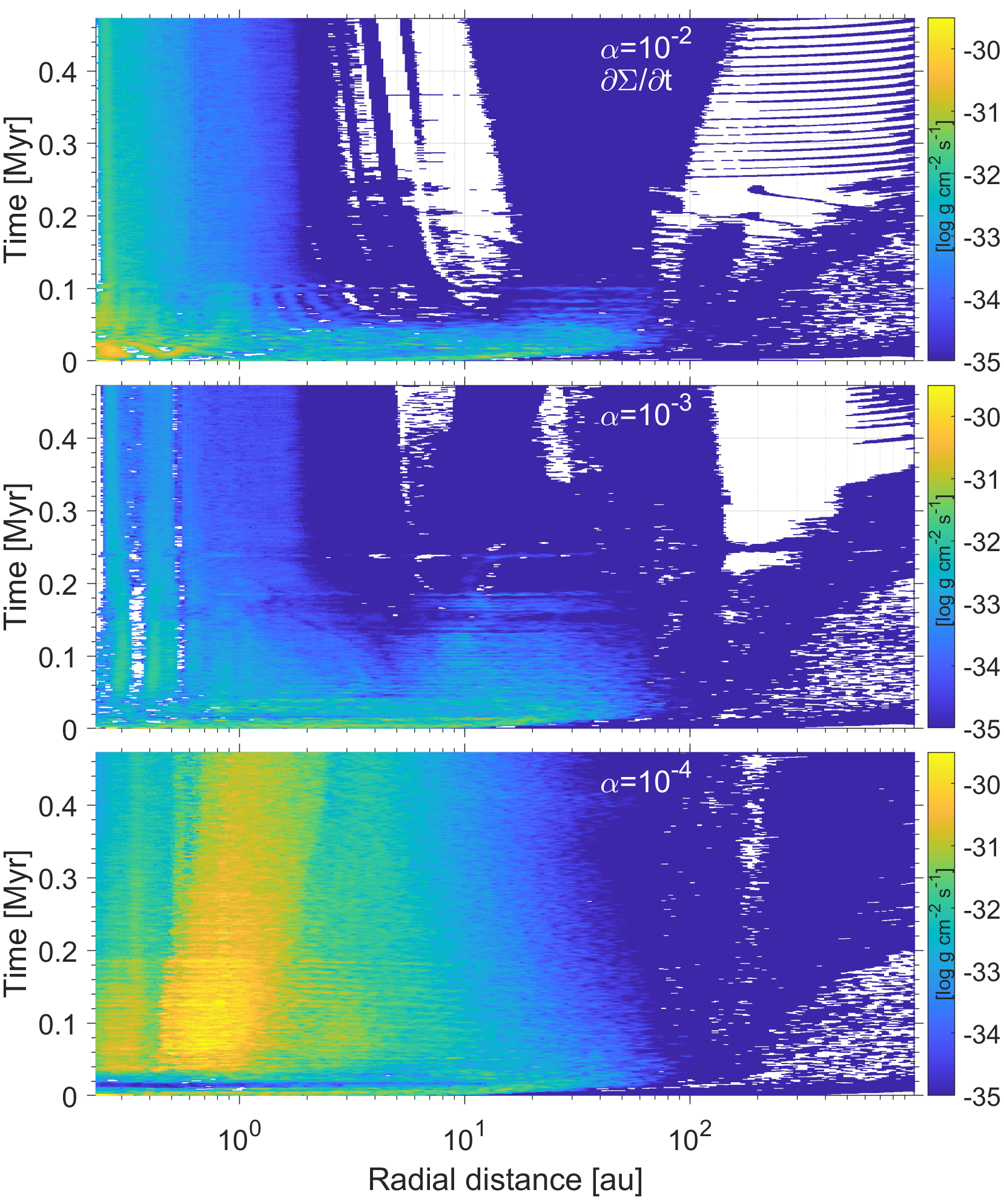}
\par\end{centering}
\caption{\label{fig:11} Space-time diagram of azimuthally averaged $\partial\Sigma/\partial t$ in our models. The top, middle, and bottom panels correspond to $\alpha=10^{-2}$, $10^{-3}$ and $10^{-4}$ models, respectively.  The white color shows the regions where $\partial\Sigma/\partial t$ is negative. }
\end{figure}

We now return to Equation~(\ref{app_dSigma_mdot}), which implies a steady-state disk with a constant density profile if $\dot{M}$ is independent of $r$ and the accumulation of matter (or increasing $\Sigma_{\rm g}$) if there is a positive radial gradient of $\dot{M}$. We calculated $\dot{M}$ using Equations~(\ref{app_mdot})--(\ref{alpha_eff}) for each computational cell and then azimuthally averaged the resulting values before calculating the radial gradients in Equation~(\ref{app_dSigma_mdot}).
The resulting space-time diagrams of  $\partial\Sigma_{\rm g}/\partial t$ for the gas disk in our models are shown in Figure~\ref{fig:11}. 
By comparing the three models, we can see the $\alpha=10^{-4}$ model demonstrates the highest (and positive) values of  $\partial\Sigma_{\rm g}/\partial t$, implying strong accumulation of matter in the disk region around 1~au, exactly where the gas and dust rings develop in Fig.~\ref{fig:2}. 
In the other two models ($\alpha=10^{-2}$ and $\alpha=10^{-3}$) the accumulation of matter is not as much pronounced and some outer disk regions even expand viscously, as indicated by the negative values of  $\partial\Sigma_{\rm g}/\partial t$.

\section{Epstein and Stokes regimes}

The common practice in numerical hydrodynamics simulations of gas-dust dynamics is to assume that the friction of dust with gas can be described in the Epstein regime when the mean free path of hydrogen molecules $\lambda$ is much longer than the size of dust grains. In this case, the friction force has a linear dependence on the difference between gas and dust velocities (see eq.~\eqref{eq:friction_force}~and~\eqref{tstop}).

\begin{figure}
\begin{centering}
\includegraphics[width=1\columnwidth]{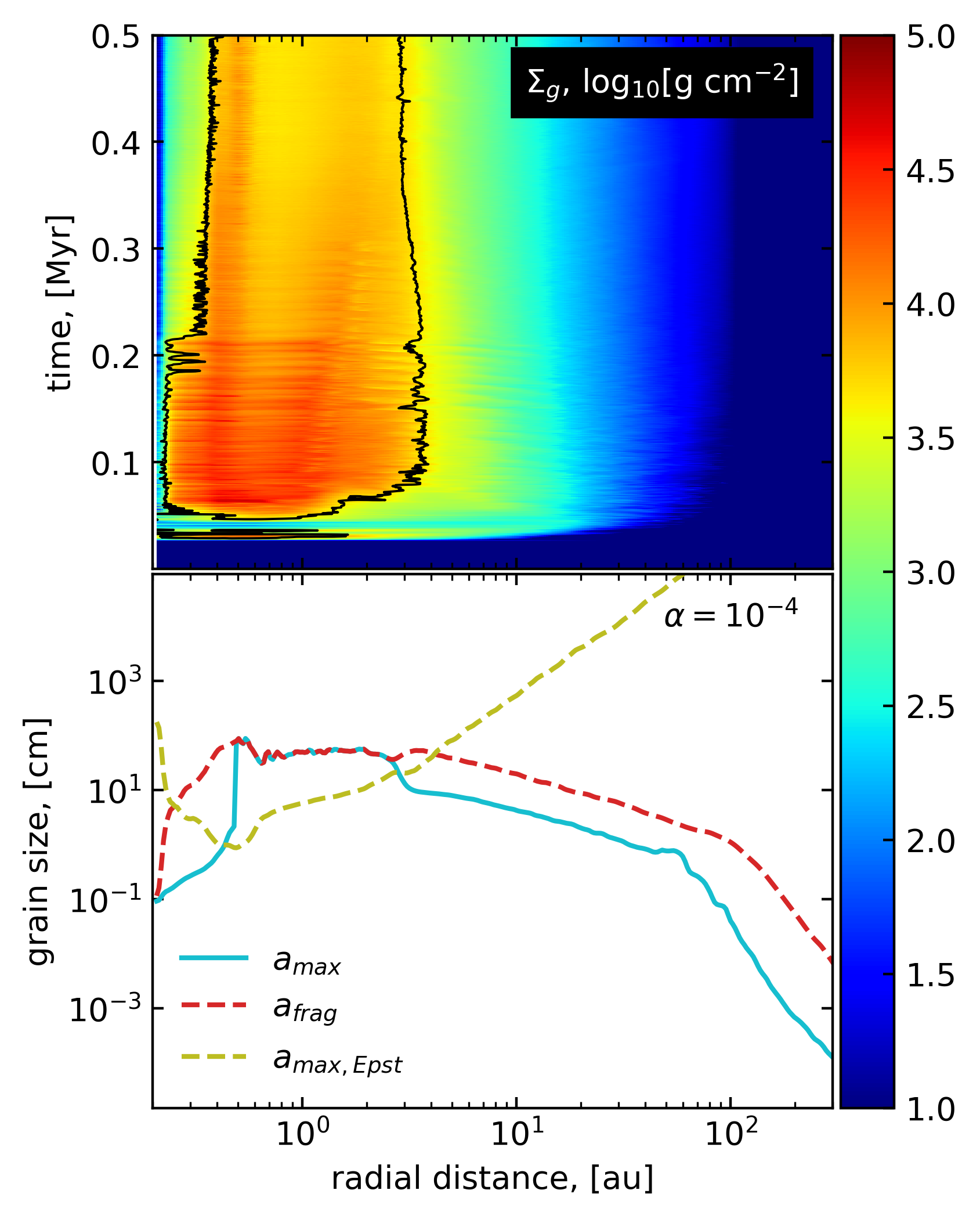}
\par\end{centering}
\caption{\label{fig:amax} {\bf Top panel:} temporal evolution of the azimuthally-averaged gas surface density in the $\alpha=10^{-4}$ model with the Stokes drag region outlined by the black curves. {\bf Bottom panel:} radial profiles of maximal dust size (cyan line), fragmentation barrier (red dashed line), and maximum size of grown dust up to which dust dynamics can be described by the Epstein drag (yellow dashed line). The profiles are shown at a time instance of $t = 500$~kyr after the onset of simulations.}
\end{figure}

However, the Epstein regime of dust dynamics can be violated either because of decreasing mean free path of H$_2$ or significant growth of dust particles. The latter is the case in the model with $\alpha = 10^{-4}$ but in the higher $\alpha$-models the Epstein regime is preserved.  The top panel of Figure~\ref{fig:amax} presents the space-time diagram of the azimuthally averaged $\Sigma_{\rm g}$ in the $\alpha=10^{-4}$ model. The disk regions where the Epstein regime is violated are outlined by the black curves. They do not extend beyond a few astronomical units, meaning that simulations with sink cells $\gg 1.0$~au may be safe to use the Epstein drag. Our models, however, have a much smaller disk inner edge, 0.2~au, and taking the Stokes drag into account becomes necessary. 

The bottom panel presents a time snapshot at the end of simulations showing the radial distribution of the azimuthally-averaged maximum dust size $a_{\rm max}$ and fragmentation barrier $a_{\rm frag}$ (see~\eqref{afrag}). In addition, we plot the maximum dust size $a_{\rm Epst}$ up to which the Epstein regime is valid, calculated as:
\begin{equation}
\label{eq:aepst}
a_{\rm Epst} = \dfrac{9}{4} \lambda, 
\end{equation}
where the mean free path of H$_2$ is found as 
\begin{equation}
\label{eq:h2freepath}
\lambda = \dfrac{m_{{\rm H}_2}}{A_{{\rm H}_2}} \dfrac{\sqrt{2\pi} H_{\rm g}}{\Sigma_{\rm g}}, 
\end{equation} 
where $m_{{\rm H}_2}$ and $A_{{\rm H}_2}$ are the mass and cross-section of the hydrogen molecule, respectively \citep{2004RiceLodato}. 
Clearly, the size of dust that is concentrated in the ring between 0.5 and 3 au is limited by the fragmentation barrier $a_{\rm frag}$. Nevertheless, the actual size remains to be high, reaching decimeters or even meters. Moreover, the actual size of dust grains $a_{\rm max}$ is systematically higher than $a_{\rm Epst}$ across the entire ring. At the inner boundary of the ring the difference between $a_{\rm Epst}$ and $a_{\rm max}$ is about two orders of magnitude. This fact indicates that taking into account the non-linear Stokes regime of dust dynamics is necessary in the $\alpha=10^{-4}$ model.

\label{Stokes}

\end{appendix}

\end{document}